\documentclass[aps,twocolumn]{revtex4-2}

\usepackage{color}
\usepackage{graphicx}
\usepackage{subfigure}
\usepackage{afterpage}
\usepackage{dcolumn}
\usepackage{bm}
\usepackage{color}
\usepackage{float}
\usepackage{amsmath}
\usepackage{bm}
\usepackage{amssymb}
\usepackage{siunitx}
\usepackage{cleveref}
\usepackage{natbib}

\def\url#1{\expandafter\string\csname #1\endcsname}
\def\bomega{\bar{\omega}}

\newcounter{defcounter}
\usepackage{amsthm}

\begin{document}

\title{Vector Edge Solitons and Domain Walls in a Nonlinear Mechanical Topological Insulator}

\author{David D. J. M. Snee}
\author{Yi-Ping Ma}
\email{yiping.ma@northumbria.ac.uk}
\affiliation{Department of Mathematics, Physics and Electrical Engineering, Northumbria University, Newcastle upon Tyne, NE1 8ST, UK}

\begin{abstract}
We report nonlinear edge waves in a 2D mechanical topological insulator. A bulk lattice consists of pendulums with on-site cubic nonlinearity connected by linear springs realizing quantum spin Hall effect. We show that the nonlinear interaction between two edge modes with equal group velocities (EGV) is described by a 1D two-component coupled nonlinear Schr\"odinger (CNLS) equation. On the interface separating two bulk lattices with opposite spin Chern numbers, we construct linear springs such that the dispersion relation exhibits EGV points with favorable CNLS coefficients. Thus, we realize nonlinear edge waves propagating along the interface, including bright-bright (BB) edge solitons for focusing CNLS coefficients, and dark-dark edge solitons, edge domain walls, and dark-bright edge solitons for defocusing CNLS coefficients. In terms of the site amplitudes, these solutions resemble bright and dark breathers. These solutions should be topologically protected when both carrier frequencies lie within a band gap, which we explicitly show by passing BB edge solitons through compact defects on the interface. We also show energy transfer in BB edge soliton collisions with potential application to collision-based computing. Generally, vector edge solitons exhibit a large parameter space for soliton collisions, which endows mechanical devices with greater potential for information processing and other functionalities.
\end{abstract}

\maketitle

\section{\label{sec:level1}Introduction}
Topological insulators (TIs) are materials which boast the unique property of conduction on the edge (surface) whilst the bulk remains insulating \cite{Hasan-Kane-2010,qi2011topological}. Although the 1D Su-Schrieffer-Heeger model is often deemed the simplest TI nowadays, the concept of TIs first entered mainstream condensed matter physics via 2D models. The integer quantum Hall (QH) effect was first observed in a landmark experiment of a 2D electron gas subject to a perpendicular magnetic field \cite{klitzing1980new}. Thouless, Kohmoto, Nightingale and den Nijs then showed theoretically that the quantized Hall conductance of the QH state results directly from an integer topological invariant of the bulk band structure called the Chern number \cite{thouless1982quantized}. The study of such TIs (Chern insulators) was initiated by Haldane, who showed that broken time-reversal symmetry is responsible for nontrivial bulk topology and chiral edge states \cite{haldane1988model}. When two materials with different bulk Chern numbers touch each other, the bulk-edge correspondence guarantees the existence of unidirectional edge modes at the interface between the two materials that are immune to backscattering in the presence of weak disorder \cite{hatsugai1993chern}. Note that one of the two materials can be the vacuum whose Chern number is zero.

Later, Kane and Mele found 2D topological phases with time-reversal symmetry preserved \cite{Kane-Mele-2005}. Such models are called quantum spin Hall (QSH) effect since they use two copies of a Chern insulator with opposite spins \cite{Bernevig-Zhang-2006}. The Chern number of either component is proportional to the spin and is often called the spin Chern number. When the spin is conserved, the two components behave independently and host the same number of edge modes propagating in opposite directions; such edge modes are called helical rather than chiral. More generally, one can allow spin-flips while preserving time-reversal symmetry. Such systems are called QSH insulators or $\mathbb{Z}_2$ TIs since they are characterized by a $\mathbb{Z}_2$ topological invariant that is the spin Chern number of either component modulo 2 \cite{kane2005z,sheng2006quantum}. At the interface between two such materials with different spin Chern numbers, the bulk-edge correspondence guarantees the existence of helical edge states that are immune to backscattering in the presence of weak time-reversal-symmetric disorder \cite{Qi-Wu-Zhang-2006-PRB}. The first realistic QSH material was predicted theoretically by Bernevig, Hughes, and Zhang \cite{bernevig2006quantum} and soon confirmed experimentally \cite{konig2007quantum}, which established TIs as a major research field.

Recently, the theoretical framework of quantum TIs was extended to photonic (electromagnetic) \cite{Ozawa-etal-2018}, cold atomic \cite{cooper2019topological}, and phononic (mechanical) \cite{shah2024colloquium} systems. Such extensions of topological phases from quantum to classical waves were initiated by Haldane and Raghu, who proposed a photonic QH analog using a 2D lattice of time-reversal symmetry-breaking elements \cite{haldane2008possible}. Their idea of using magneto-optic effect was realized experimentally in the microwave domain \cite{Wang-etal-2009}, but alternative approaches were needed in the optical domain. A pioneering experiment uses Floquet engineering to realize the Haldane model for QH effect, marking the birth of photonic Floquet TIs \cite{Rechtsman-etal-2013}. Meanwhile, pioneering work on photonic QSH analogs constructs pseudo-spins in an artificial magnetic field on various platforms including bianisotropic metamaterials \cite{Khanikaev-etal-2013} and coupled silicon ring resonators \cite{hafezi2013imaging}. Since these early discoveries, the research field of topological photonics has grown rapidly \cite{Ozawa-etal-2018}.

In mechanical systems, topological edge modes can be realized at either zero frequency or high frequencies \cite{Huber-2016-NP}. Here we focus on high-frequency topological mechanics, which permits topological phonon transport \cite{shah2024colloquium} and thus parallels topological photonics. In earliest proposals of phononic QH analogs, time-reversal symmetry breaking is first conjectured for microtubules in biophysics \cite{prodan2009topological}, but it is first realized in optomechanical crystals \cite{peano2015topological}. Pioneering work on phononic Chern insulators achieves intrinsic time-reversal symmetry breaking using coupled gyroscopes \cite{NKRVTI_PNAS_2015,wang2015topological} or Coriolis force \cite{kariyado2015manipulation,wang2015coriolis}. Meanwhile, a pioneering experiment uses coupled pendula to realize a mechanical QSH analog, marking the birth of mechanical TIs (MTIs) \cite{Susstrunk-Huber-2015}. Topological phonon transport is actively studied in not only discrete systems as reviewed above, but also acoustic systems (gases or liquids) and elastic systems (solids) on many length scales \cite{shah2024colloquium}.

Although classical TIs including photonic, cold atomic, and mechanical TIs resemble quantum TIs linearly, the former can also exhibit rich nonlinear phenomena absent in the latter. The emerging field of nonlinear topological photonics combines topology and nonlinearity to achieve advanced functionalities in optical devices \cite{smirnova2020nonlinear,ablowitz2022nonlinear,szameit2024discrete}. A salient feature of nonlinear TIs is the possible existence of localized states on the edge called edge solitons. In 2D TIs, an edge soliton with a narrow spectral (wide spatial) envelope along the edge is topologically protected if its carrier frequency lies within the topological band gap. Such continuous edge solitons are extensively studied in photonic Floquet TIs \cite{Ablowitz-Curtis-Ma-2014,Ablowitz-Ma-2015,ablowitz2015adiabatic,lumer2016instability,ablowitz2017tight,ivanov2020edge,ivanov2020bragg,bazhan2021josephson,shi2021topological,ren2023floquet} and polariton TIs \cite{kartashov2016modulational,gulevich2017exploring,li2018lieb,zhang2019interface}. Meanwhile, discrete edge solitons in photonic Floquet TIs have also attracted great interest \cite{Leykam-Chong-2016}. For bright edge solitons, there are typically a two-parameter continuous family with low amplitudes and a one-parameter discrete family with arbitrary amplitudes. Such scalar edge solitons in photonic TIs are recently observed in pioneering experiments \cite{mukherjee2020observation,mukherjee2021observation,zhang2020observation,kartashov2022observation}, paving the way for device applications. Meanwhile, vector edge solitons in photonic or cold atomic TIs are also found on the interface between a TI and its partner \cite{ivanov2020vector}, the edge of a TI with two band gaps \cite{ivanov2021topological}, or a combination thereof \cite{tian2023vector}. Generally, vector solitons have a larger parameter space and thus exhibit richer dynamics than their scalar counterparts.

By contrast, nonlinear topological phononics is still in its early stages, but some central themes have emerged. In Ref.~\cite{Pal-etal-2018}, amplitude-dependent edge modes are found in 1D and 2D MTIs with inter-site cubic nonlinearity. In the 2D case, the MTI studied is a bi-layered lattice realizing a mechanical QSH analog \cite{Pal-etal-2016}, and high-amplitude edge waves are found numerically. In Ref.~\cite{snee2019edge}, scalar edge solitons are found analytically in a 2D MTI with on-site cubic nonlinearity that describes the first experimental realization of a mechanical QSH analog \cite{Susstrunk-Huber-2015}. Such edge solitons are low-amplitude edge waves preserving their shapes during propagation and thus parallel those in 2D nonlinear photonic or cold atomic TIs. Meanwhile, there are proposals to create topologically protected edge channels using nonlinearity in a 1D lattice with inter-site cubic nonlinearity with alternating signs \cite{chaunsali2019self} or a 2D lattice with on-site cubic nonlinearity via zone folding \cite{darabi2019tunable}. Recently, the existence, stability, and dynamics of nonlinear topological edge states are extensively studied in 1D MTIs with on-site cubic nonlinearity \cite{chaunsali2021stability,many2022nonlinear}, inter-site cubic nonlinearity \cite{tempelman2021topological,rosa2023amplitude} possibly with local resonators \cite{legrande2024introduction}, or on-site sine nonlinearity \cite{ezawa2022nonlinear}. However, nonlinear topological edge states in 2D MTIs are rarely explored to our knowledge. Specifically, although scalar edge solitons are reported \cite{snee2019edge}, vector edge solitons remain open.

In this paper, we construct a 2D nonlinear MTI that exhibits vector edge solitons and domain walls (DWs). In Section \ref{sec:ml-dr}, we review the design of a 2D MTI that is a mechanical QSH analog. In Sections \ref{sec:cnls}, we derive a 1D coupled nonlinear Schr\"odinger (CNLS) equation from a 2D MTI with on-site cubic nonlinearity and review key solutions of the CNLS equation including vector solitons and DWs. In Sections \ref{sec:mti-2} \& \ref{sec:ves-dw}, we first design a 2D nonlinear MTI with two topological sectors that provides a flexible platform to access different regimes of the 1D CNLS equation. Then, we study numerically the propagation of vector edge solitons and DWs in this system. In Sections \ref{sec:tp-bb} \& \ref{sec:col-bb}, we further explore bright-bright (BB) edge solitons including their robust propagation around compact defects and their collision properties with possible applications to computing. The paper concludes in Section \ref{sec:disc} with some directions for future research.

\section{Mechanical topological insulator and dispersion relation}\label{sec:ml-dr}
Our 2D MTI is an adaptation of the first mechanical analogue of a QSH insulator realized experimentally in Ref.~\cite{Susstrunk-Huber-2015}. The mechanical lattice consists of a collection of pendula connected by linear springs. The QSH effect uses two copies of the Hofstadter model \cite{hofstadter1976energy} on a square lattice indexed by $(r,s)$. Thus, the Hamiltonian is
\begin{equation*}
    \hat{H}=\sum_{\alpha=\pm}\hat{H}_\alpha,
\end{equation*}
where the Hamiltonian for each spin is given by
\begin{equation*}
        \hat{H}_\alpha=f_0\sum_{r,s}\left(\hat{a}_{r,s,\alpha}^\dagger\hat{a}_{r,s+1,\alpha}
    +e^{i\alpha\Phi s}\hat{a}_{r,s,\alpha}^\dagger\hat{a}_{r+1,s,\alpha}+\mathrm{H.c.}\right).
\end{equation*}
Here, $\alpha$ is the spin-index, $f_0$ is the hopping amplitude, $\hat{a}_{r,s,\alpha}^\dagger$ and $\hat{a}_{r,s,\alpha}$ are respectively the creation and annihilation operators of a particle with spin $\alpha$ at site $(r,s)$, $\Phi$ is the magnetic flux, and H.c.~denotes Hermitian conjugacy. The choice $\Phi=2\pi/3$ makes $\hat{H}$ periodic on a $1\times3$ unit cell, so the Hamiltonian matrix is
\begin{equation}\label{eq:MTI_Hamiltionian_matrix}
    H=\left(\begin{array}{cc}
    H_+ & 0\\
    0 & H_-
    \end{array}\right),
\end{equation}
where $H_{\pm}$ are $3 \times 3$ matrices.

The key insight of Ref.~\cite{Susstrunk-Huber-2015} is that the Hamiltonian matrix $H$ can be made real symmetric via a similarity transform. The resulting matrix can then be made positive-definite and serve as the dynamical matrix for coupled oscillators. In the experimental setup, each site of the square lattice hosts two 1D pendula $(x_{r,s},y_{r,s})$, both of which swing only in the $s$ direction, and the linear connections between neighboring sites are realized using springs possibly with lever arms. Hereafter, we group the lattice sites into unit cells indexed by $(r,S)$ with each cell consisting of 3 sites $(x^{(j)}_{r,S},y^{(j)}_{r,S})$ for $j=0,1,2$.

To enable nonlinear waves in this MTI, we account for the inherent cubic (Duffing) nonlinearity of the pendula. The equations of motion for the 6 pendula $(x^{(j)},y^{(j)})$, $j=0,1,2$, in the unit cell $(r,S)$ are written explicitly as
\begin{align}
\ddot{x}^{(0)}_{r,S}(t) =&-(\omega_0^2+A_sf)x^{(0)}_{r,S}+\sigma (x^{(0)}_{r,S})^3 \notag \\ & +f(x^{(1)}_{r,S}+x^{(2)}_{r,S-1}+x^{(0)}_{r+1,S}+x^{(0)}_{r-1,S}),  \label{eq:eqx0} \\
\ddot{y}^{(0)}_{r,S}(t) =&-(\omega_0^2+A_sf)y^{(0)}_{r,S}+\sigma (y^{(0)}_{r,S})^3 \notag \\ & +f(y^{(1)}_{r,S}+y^{(2)}_{r,S-1}+y^{(0)}_{r+1,S}+y^{(0)}_{r-1,S}),   \label{eq:eqy0} \\
\ddot{x}^{(1)}_{r,S}(t) =&-(\omega_0^2+A_sf)x^{(1)}_{r,S}+\sigma (x^{(1)}_{r,S})^3 \notag \\ & +f(x^{(0)}_{r,S}+x^{(2)}_{r,S})-\frac{f}{2}(x^{(1)}_{r+1,S}+x^{(1)}_{r-1,S}) \notag \\ & +\frac{\sqrt{3}f}{2}(y^{(1)}_{r+1,S}-y^{(1)}_{r-1,S}), \label{eq:eqx1} \\ 
\ddot{y}^{(1)}_{r,S}(t) =&-(\omega_0^2+A_sf)y^{(1)}_{r,S}+\sigma (y^{(1)}_{r,S})^3 \notag \\ & +f(y^{(0)}_{r,S}+y^{(2)}_{r,S})-\frac{f}{2}(y^{(1)}_{r+1,S}+y^{(1)}_{r-1,S}) \notag \\ & +\frac{\sqrt{3}f}{2}(-x^{(1)}_{r+1,S}+x^{(1)}_{r-1,S}),  \label{eq:eqy1} \\
\ddot{x}^{(2)}_{r,S}(t)  =&-(\omega_0^2+A_sf)x^{(2)}_{r,S}+\sigma (x^{(2)}_{r,S})^3 \notag \\ & +f(x^{(0)}_{r,S+1}+x^{(1)}_{r,S})-\frac{f}{2}(x^{(2)}_{r+1,S}+x^{(2)}_{r-1,S}) \notag \\ & +\frac{\sqrt{3}f}{2}(-y^{(2)}_{r+1,S}+y^{(2)}_{r-1,S}),  \label{eq:eqx2} \\
\ddot{y}^{(2)}_{r,S}(t)  =&-(\omega_0^2+A_sf)y^{(2)}_{r,S}+\sigma (y^{(2)}_{r,S})^3 \notag \\ & +f(y^{(0)}_{r,S+1}+y^{(1)}_{r,S})-\frac{f}{2}(y^{(2)}_{r+1,S}+y^{(2)}_{r-1,S}) \notag \\ & +\frac{\sqrt{3}f}{2}(x^{(2)}_{r+1,S}-x^{(2)}_{r-1,S}).  \label{eq:eqy2}
\end{align}
Here, $t$ denotes time, $\cdot$ denotes time derivative, $f$ describes the linear restoring forces of the springs, and the pendula are assumed identical with angular frequency $\omega_0$. Consistent with Refs.~\cite{Susstrunk-Huber-2015,snee2019edge}, we choose $f=4.16\pi^2$ and $\omega_0=3\pi/2$. The nonlinear coefficient must be $\sigma=\omega_0^2/6$ to yield a cubic approximation to the sinusoidal restoring force of a pendulum. The self-coupling coefficient $A_s$ depends on the detailed setup as explained next.

\begin{figure}
    \centering
    \includegraphics[width=.4\textwidth]{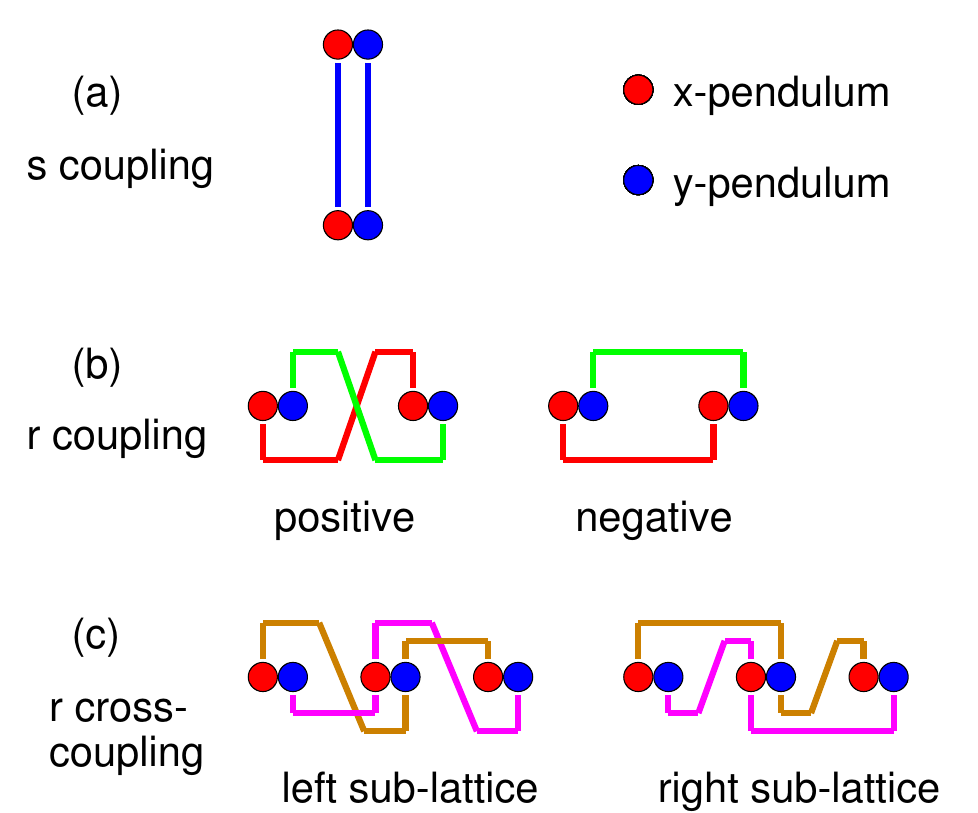}
    \caption{(Color online) Schematic view of the connections between neighboring pendula in the 2D MTI. The $r$-direction is horizontal, the $s$-direction is vertical, and all pendula swing in the $s$-direction. (a) Simple springs connecting $x$--$x$ and $y$--$y$ pendula in the $s$-direction. (b) Complex springs connecting $x$--$x$ and $y$--$y$ pendula in the $r$-direction. (c) Complex springs connecting $x$--$y$ pendula in the $r$-direction. Here, a complex spring with two springs and one lever arm realizes a negative coupling, while a complex spring with three springs and two lever arms realizes a positive coupling; see Ref.~\cite{Susstrunk-Huber-2015} for a physical depiction. If the magnetic flux $\Phi$ changes sign, then the $r$ cross-couplings in panel (c) change sign. As explained in Section \ref{sec:mti-2}, these two types of $r$ cross-couplings define the two sub-lattices of the 2D MTI with two topological sectors.}
    \label{fig:lattice_connections}
\end{figure}

For a simple spring connecting a pendulum and its neighbor, the restoring force, which is proportional to the relative displacement, yields a positive coupling to the neighbor and a negative self-coupling. As shown in the first two terms of the second rows in Eqs.~(\ref{eq:eqx0}--\ref{eq:eqy2}), all couplings to the neighbor in the $s$-direction are $f>0$ and thus can be realized by simple springs; see Fig.~\ref{fig:lattice_connections}(a).

In contrast, any connection in the $r$-direction cannot be realized by a simple spring since the displacements are in the $s$-direction. Thus, lever arm(s) must be used to rotate the displacements around pivot(s). As shown in the last two terms of the second rows in Eqs.~(\ref{eq:eqx0}--\ref{eq:eqy2}), the couplings to the neighbor in the $r$-direction are $f$ for $(x^{(0)},y^{(0)})$ and $f\cos\Phi=-f/2$ for $(x^{(1)},y^{(1)})$ and $(x^{(2)},y^{(2)})$. The former requires two lever arms, while the latter requires one lever arm; see Fig. \ref{fig:lattice_connections}(b).

The above two types of couplings, i.e., $s$ couplings and $r$ couplings, connect either $x$--$x$ or $y$--$y$ pendula. Meanwhile, as shown in the third rows in Eqs.~(\ref{eq:eqx1}--\ref{eq:eqy2}), a third type of couplings connect $x$--$y$ pendula in the $r$-direction for $(x^{(1)},y^{(1)})$ and $(x^{(2)},y^{(2)})$. These $r$ cross-couplings are $f\sin\Phi=\sqrt{3}f/2$ times alternating signs, so they change sign when $\Phi$ changes sign; see Fig.~\ref{fig:lattice_connections}(c).

Besides a coupling to the neighbor, either a simple spring or a complex spring with lever arm(s) yields a negative self-coupling. Thus, the total self-coupling for $(x^{(1)},y^{(1)})$ and $(x^{(2)},y^{(2)})$ is $-A_sf$ where $A_s=3+\sqrt{3}$. The total self-coupling for $(x^{(0)},y^{(0)})$ is $-4f$ but can be made $-A_sf$ using springs attached to walls.

For ease of computation, the nonlinear equations of motion (\ref{eq:eqx0}--\ref{eq:eqy2}) may be written in the compact matrix form
\begin{equation}\label{eq:eqn-of-motion}
\boldsymbol{\ddot{X}}_{r,S}(t) = (\boldsymbol{\mathcal{L}}\boldsymbol{X})_{r,S} + \sigma \boldsymbol{\mathcal{N}}_{r,S}
\end{equation}
where $\boldsymbol{X}=[x^{(0)}, y^{(0)}, x^{(1)}, y^{(1)}, x^{(2)}, y^{(2)}]^{T}$, $\boldsymbol{\mathcal{L}}$ is the matrix encoding the linear couplings, and $\boldsymbol{\mathcal{N}}=\boldsymbol{X}^3$ is the cubic nonlinearity. In the linear limit $\sigma=0$, this system fits into the classification scheme of topological phonons \cite{susstrunk2016_PNAS}. Since velocity-dependent forces are absent, this system belongs to the category of reciprocal metamaterials.

To obtain all possible symmetries of this system, we study the dynamical matrix $D(\boldsymbol{k})$ with 2D wavevector $\boldsymbol{k}$, which is the 2D Fourier transform of $\boldsymbol{\cal L}$. As shown in Ref.~\cite{susstrunk2016_PNAS}, the structure of $D(\boldsymbol{k})$ exhibits a ${\cal T}$ symmetry that squares to $+1$. Moreover, this symmetry can be augmented to a ${\cal T}^*$ symmetry that squares to $-1$. Here, the ${\cal T}$ symmetry generalizes the time-reversal symmetry in quantum TIs but does not imply the reversal of time in the mechanical setting. Overall, this system belongs to symmetry class AII in 2D with the presence of the ${\cal T}$ and ${\cal T}^*$ symmetries and the absence of other symmetries, and the bulk topological index can be shown to be $\mathbb{Z}_2$; see Ref.~\cite{susstrunk2016_PNAS} for detailed analyses of this and related systems.

\begin{figure}
\centering
\includegraphics[width=.45\textwidth]{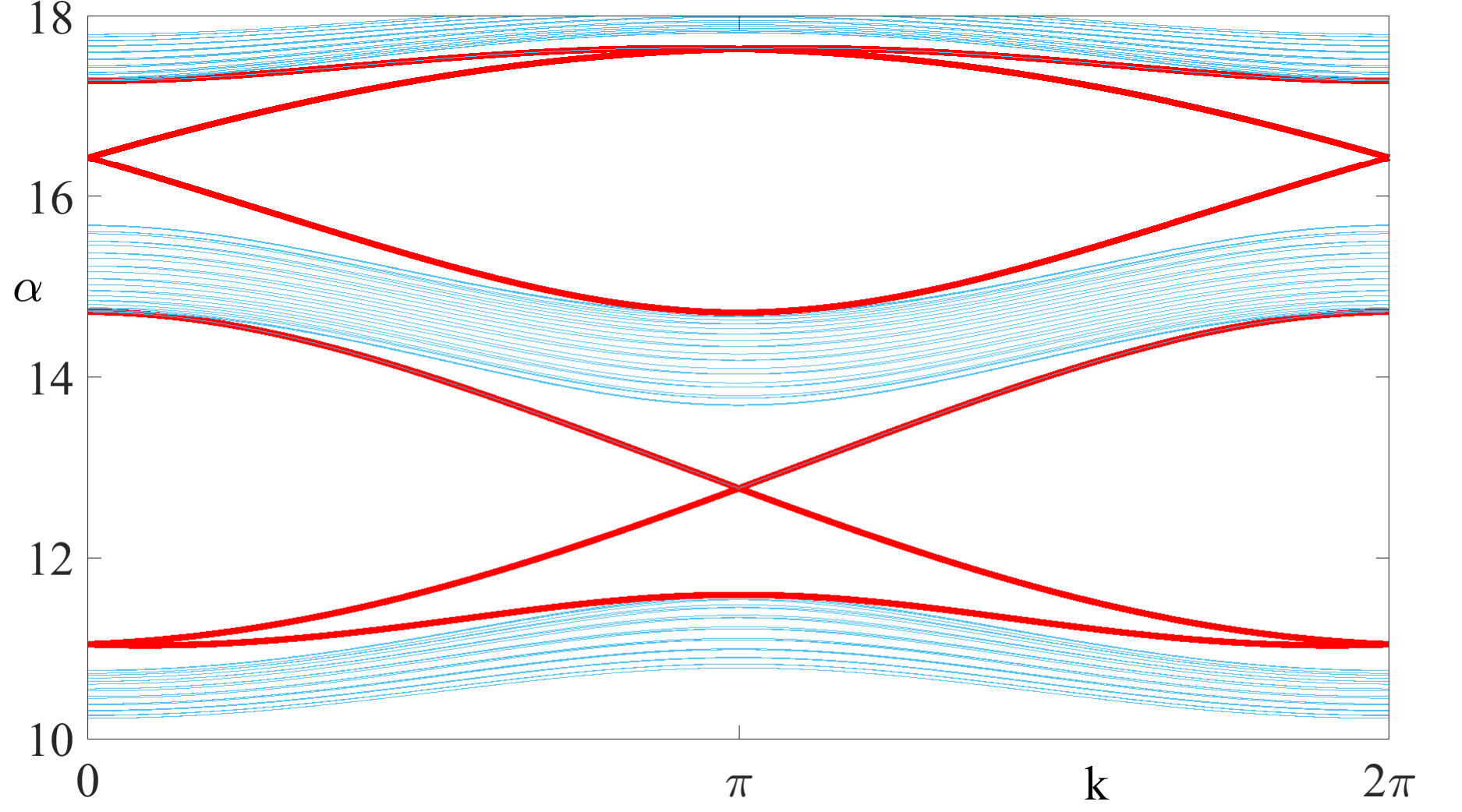}
    \caption{(Color online) The dispersion relation $\alpha(k)$ obtained from numerical solutions of the eigenvalue problem in Eq.~(\ref{eq:linear_problem}). The bulk and edge spectra are shown in blue and red respectively. This dispersion relation does not change when the magnetic flux $\Phi$ changes sign.}
    \label{fig:dispersion_relation}
\end{figure}

Due to the bulk-edge correspondence, the nontrivial topology of the bulk band structure guarantees the existence of topologically protected helical edge states at the interface between this 2D MTI and the vacuum. To find the dispersion relation of edge states along any direction, say $S$, we consider the linear problem, i.e., Eq.~(\ref{eq:eqn-of-motion}) with $\sigma=0$, and apply the 1D Fourier transform $\boldsymbol{X}_{r,S}=e^{i(Sk-t\alpha(k))}\boldsymbol{X}_{r} + c.c.$ where $k$ is the wavenumber in $S$, $\alpha(k)$ is the dispersion relation, and $c.c.$ denotes complex conjugate. This yields the eigenvalue problem
\begin{equation}\label{eq:linear_problem}
    \mathcal{L}(k)\boldsymbol{X}_{r}=-\alpha(k)^2\boldsymbol{X}_{r} 
\end{equation}
where $\mathcal{L}(k)$ denotes the matrix $\boldsymbol{\cal L}$ in Eq.~(\ref{eq:eqn-of-motion}) after the 1D Fourier transform, and the eigenvector $\boldsymbol{X}_{r}$ is normalized such that $\|\boldsymbol{X}_r\|^2_2=\sum_j |(\boldsymbol{X}_r)_{j}|^2=1$.

The dispersion relation $\alpha(k)$, $k \in [0,2\pi)$, can be numerically computed on a finite 1D domain in $r$. Figure \ref{fig:dispersion_relation} shows this dispersion relation for $N_r=30$ sites with the bulk (continuous) spectrum shown in blue and the edge (discrete) spectrum shown in red. The ${\cal T}$ symmetry implies that the band structures, both bulk and edge, are symmetric with respect to $k=\pi$ \cite{susstrunk2016_PNAS}. This generalized Kramers' theorem implies that any eigenvector at $(k,\alpha)$ has a Kramers partner at $(2\pi-k,\alpha)$, and together they form a Kramers pair. Since the unit cell has three sites, there are three bulk bands and two band gaps. For any $\alpha$ in a band gap, there is a Kramers pair of topologically protected edge states. This pair of edge states are called helical since they have opposite group velocities.

\section{Derivation and solutions of coupled nonlinear Schr\"odinger equation}\label{sec:cnls}

Consider a generic 2D nonlinear MTI with two branches of the dispersion relation, $\alpha(k)$ and $\beta(k)$. To find weakly nonlinear solutions, we let the multiple scale ansatz be a linear superposition of two edge modes:
\begin{align}\label{eq:CM_ansatz}
\boldsymbol{X}_{r,S}(t)=\epsilon\big\{\mathcal{A}(&\tilde{S},\tau)e^{i(Sk_0-t\alpha_0)}\boldsymbol{X}_r^{(1)}+ \notag \\ &\mathcal{B}(\tilde{S},\tau)e^{i(Sk_0-t\beta_0)}\boldsymbol{X}_r^{(2)}+c.c.\big\} + O(\epsilon^2),
\end{align}
where the small parameter $0<\epsilon \ll 1$ is the amplitude, $k_0$ is the carrier wavenumber, and $\alpha_0\equiv\alpha(k_0)$ and $\beta_0\equiv\beta(k_0)$ are the two carrier frequencies. The two edge states $\boldsymbol{X}_r^{(\cdot)}$ are eigenvectors of ${\cal L}(k_0)$ and normalized such that $\|\boldsymbol{X}_r^{(1)}\|^2_2=\|\boldsymbol{X}_r^{(2)}\|^2_2=1$. The spectral envelope width is assumed to be $\epsilon$, so the two scalar envelopes $\mathcal{A}$ and $\mathcal{B}$ depend on the slow space variable $\tilde{S}\equiv\epsilon( S - V_g t)$, where $V_g$ is the group velocity. Moreover, both envelopes are assumed to evolve in the slow time variable $\tau\equiv\epsilon^2 t$.

Substituting the ansatz (\ref{eq:CM_ansatz}) into Eq.~(\ref{eq:eqn-of-motion}) and expanding in powers of $\epsilon$, the $O(\epsilon)$ and $O(\epsilon^2)$ equations are trivial when $V_g=\alpha'_0=\beta'_0$, where $\alpha'_0\equiv\alpha'(k_0)$ and $\beta'_0\equiv\beta'(k_0)$. Generically, this equal group velocities (EGV) condition $\alpha'_0=\beta'_0$ defines a discrete set of $k_0$'s. At $O(\epsilon^3)$, one takes the inner product of the $\mathcal{A}$ equation with $\boldsymbol{X}_r^{(1)}$ and the inner product of the $\mathcal{B}$ equation with $\boldsymbol{X}_r^{(2)}$, with the inner product defined as $\langle\boldsymbol{g},\boldsymbol{h} \rangle=\sum_j g_j^*h_j$, to yield the 1D CNLS equation:
\begin{equation}\label{eq:MTI_CNLS_EQN1}
\begin{aligned}
&i\mathcal{A}_{\tau} + \frac{\alpha_0''}{2}\mathcal{A}_{\tilde{S}\tilde{S}} + 3\mathcal{A}(\tilde{\sigma}_1 |\mathcal{A}|^2 +  2\tilde{\sigma}_2|\mathcal{B}|^2) = 0, \\
&i\mathcal{B}_{\tau} + \frac{\beta_0''}{2}\mathcal{B}_{\tilde{S}\tilde{S}} + 3\mathcal{B}(\tilde{\sigma}_3 |\mathcal{B}|^2 +  2\tilde{\sigma}_4|\mathcal{A}|^2)  = 0,
\end{aligned}
\end{equation}
where $\alpha_0''\equiv\alpha''(k_0)$, $\beta_0''\equiv\beta''(k_0)$, and
\begin{align*}
&\tilde{\sigma}_1=\frac{\sigma}{2\alpha_0} \|\boldsymbol{X}_r^{(1)}\|^4_4, \quad \tilde{\sigma}_2=\frac{\sigma}{2\alpha_0}\|\boldsymbol{X}_r^{(1)}\boldsymbol{X}_r^{(2)}\|^2_2, \\ &\tilde{\sigma}_3=\frac{\sigma}{2\beta_0} \|\boldsymbol{X}_r^{(2)}\|^4_4, \quad \tilde{\sigma}_4=\frac{\sigma}{2\beta_0}\|\boldsymbol{X}_r^{(1)}\boldsymbol{X}_r^{(2)}\|^2_2.
\end{align*}
Note that setting ${\cal A}=0$ or ${\cal B}=0$ in Eq.~\eqref{eq:MTI_CNLS_EQN1} recovers the scalar nonlinear Schr\"odinger (NLS) equation in Ref.~\cite{snee2019edge}.

The two-component CNLS equation is the universal envelope equation for nonlinear interactions between two quasi-monochromatic plane waves \cite{manakov1974theory,zakharov1975theory}. Over decades, this equation is widely applied to optics \cite{menyuk1987nonlinear,chen1997coupled,ostrovskaya1999interaction}, mechanics \cite{shukla2006instability,ablowitz2015interacting,griffiths2006modulational}, Bose-Einstein condensates (BECs) \cite{pitaevskii2016bose,kevrekidis2015defocusing}, and other fields. A key feature of this equation is its complete integrability for certain coefficients \cite{zakharov1982integrability,sahadevan1986painleve,wang2010integrable}. For Eq.~\eqref{eq:MTI_CNLS_EQN1}, the integrability condition consists of three equalities respectively between the dispersion, $|{\cal A}|^2$, and $|{\cal B}|^2$ coefficients. Thus, local edge integrability at a single $k_0$ requires four conditions:
\begin{enumerate}
\item EGV: $\alpha'_0=\beta'_0$;
\item Curvature condition: $\alpha_0''=\beta_0''$;
\item Angle condition: $\tilde{\sigma}_1\tilde{\sigma}_3=4\tilde{\sigma}_2\tilde{\sigma}_4$, i.e., $\|\boldsymbol{X}_r^{(1)}\|_4^2\|\boldsymbol{X}_r^{(2)}\|_4^2=2\|\boldsymbol{X}_r^{(1)}\boldsymbol{X}_r^{(2)}\|_2^2$, implying that the angle between the two vectors $|\boldsymbol{X}_r^{(1)}|^2$ and $|\boldsymbol{X}_r^{(2)}|^2$ is $\pi/3$;
\item Final condition: $\tilde{\sigma}_1=2\tilde{\sigma}_4$, i.e., $\beta_0/\alpha_0=2\|\boldsymbol{X}_r^{(1)}\boldsymbol{X}_r^{(2)}\|_2^2/\|\boldsymbol{X}_r^{(1)}\|_4^4$.
\end{enumerate}
To satisfy all four conditions, three free parameters are needed in addition to $k_0$. If we further require global edge integrability for all $k_0$, then EGV for all $k_0$ is equivalent to the shift condition $\alpha(k_0)=\beta(k_0)+\text{constant}$ for all $k_0$, so the curvature condition is redundant, but the angle and final conditions are still needed for all $k_0$. Thus, edge integrability is theoretically possible but practically hard, so it will be pursued elsewhere.

Many studies in the non-integrable regime focus on the symmetric CNLS equation defined by three equalities respectively between the dispersion, self phase modulation (SPM), and cross phase modulation (XPM) coefficients. This equation can inherit key dynamical features from its integrable limit known as the Manakov system \cite{manakov1974theory}. These include BB solitons in the focusing regime \cite{yang1997classification,yang2000fractal,smyth2001radiative} and dark-bright (DB) and dark-dark (DD) solitons in the defocusing regime \cite{haelterman1994bifurcations,sheppard1997polarized}. By contrast, DWs can form in the symmetric CNLS equation in the defocusing regime \cite{haelterman1994polarization,haelterman1994vector,malomed1994optical}, but not in the integrable limit. In the degenerate case $\alpha(k)=\beta(k)$ for $k\approx k_0$ and $|\boldsymbol{X}_r^{(1)}|^2=|\boldsymbol{X}_r^{(2)}|^2$, our Eq.~\eqref{eq:MTI_CNLS_EQN1} is indeed symmetric, and the XPM coefficient is twice the SPM coefficient.

As shown below, our MTI exhibits near degeneracy, so Eq.~\eqref{eq:MTI_CNLS_EQN1} is nearly symmetric. Localized solutions to the symmetric CNLS equation reviewed above persist into this regime, but we do not assume near symmetry to find such solutions. Instead, we follow a general program to classify localized solutions to the CNLS equation without restricting the signs or magnitudes of any coefficients \cite{snee2024domain}; see Appendix \ref{app:cnls} for a review. A key conclusion is that the parameter space for vector soliton collisions is 5D: the first three parameters are the group velocity and the respective frequencies of the two components, while the remaining two parameters are the wavenumber for a dark component and the phase (polarization) for a bright component. This endows vector solitons with greater potential for information processing than scalar solitons.

\section{Mechanical topological insulator with two topological sectors}\label{sec:mti-2}

\begin{figure}
\centering
\includegraphics[width=.4\textwidth]{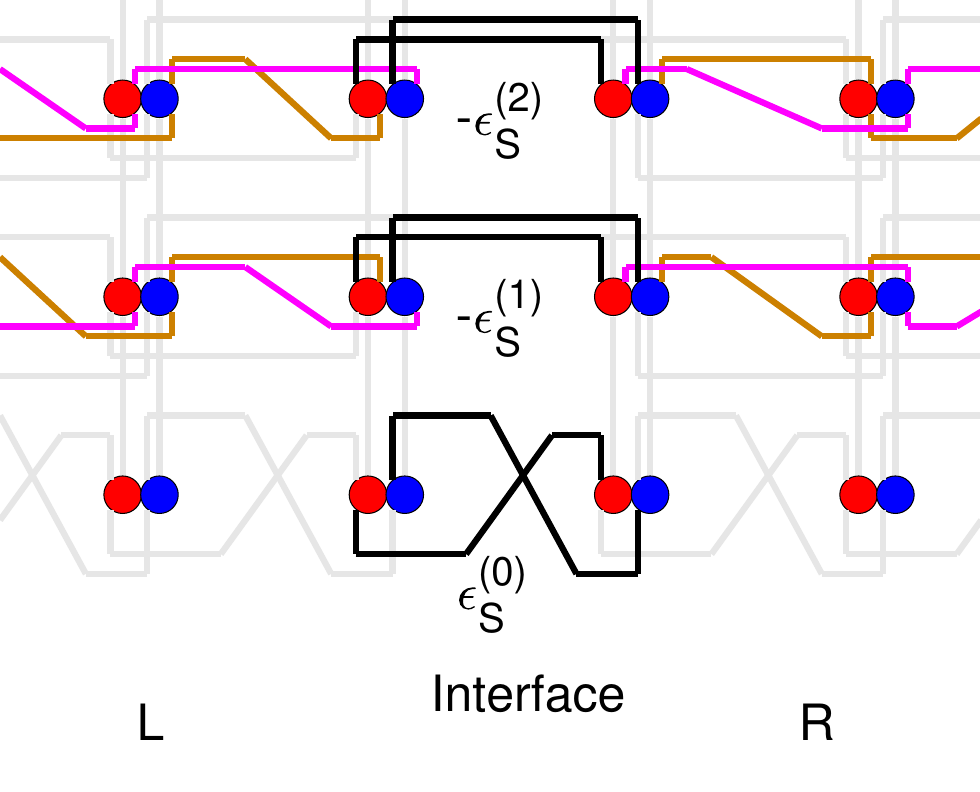}
    \caption{(Color online) MTI with two topological sectors. The two sub-lattices are connected by an interface. As in Fig.~\ref{fig:lattice_connections}, red (blue) points represent $x$ ($y$) pendula, the $r$-direction is horizontal, and the $s$-direction is vertical. Bottom row represents $(x_{r,S}^{(0)},y_{r,S}^{(0)})$ sites, middle row $(x_{r,S}^{(1)},y_{r,S}^{(1)})$, and top row $(x_{r,S}^{(2)},y_{r,S}^{(2)})$. Positive (negative) couplings are realized by two (one) lever arm(s). Brown and magenta cross-couplings are reversed in the left (L) and right (R) sub-lattices, while any unchanged couplings are grayed out. Black couplings form the interface between the two sub-lattices.}
    \label{fig:lattice_interface}
\end{figure}

Our MTI must be designed to exhibit an EGV point in the dispersion relation, i.e., a $k_0$ satisfying the EGV condition. In photonic or cold atomic TIs, an EGV point can be found on either the interface between a TI and its partner \cite{ivanov2020vector} or the edge of a TI with two band gaps \cite{ivanov2021topological}. The MTI in Section \ref{sec:ml-dr} does have two band gaps, but it is hard to modify its edge to both find an EGV point and enable favorable CNLS coefficients. Thus, we construct an interface between the MTI and its partner similarly to Ref.~\cite{ivanov2020vector}. Hereafter, the MTI always refers to this system of two sub-lattices connected by an interface.

We construct the two sub-lattices, left and right, such that the left sub-lattice has flux $\Phi=2\pi/3$ as in Section \ref{sec:ml-dr}, and the right sub-lattice has the opposite flux $\Phi=-2\pi/3$. This action of flipping the flux is equivalent to flipping the subscripts of the block matrices in Eq.~(\ref{eq:MTI_Hamiltionian_matrix}), such that the diagonal blocks are now $[H_-,H_+]$. This action is also equivalent to flipping the signs of all cross-couplings between $x$--$y$ pendula in the equations of motion (\ref{eq:eqx1}--\ref{eq:eqy2}), while leaving all couplings between $x$--$x$ and $y$--$y$ pendula unchanged. Practically, for all cross-couplings in Eqs.~(\ref{eq:eqx1}--\ref{eq:eqy2}), the number of lever arms must be changed from one to two, and vice versa. This difference between the two sub-lattices is shown in Fig.~\ref{fig:lattice_interface}, where the brown/magenta paths show the differences in the number of lever arms used for all cross-couplings. The two sub-lattices represent two topological sectors because their spin Chern numbers are opposite. This setup was initially used to show topological protection in Ref.~\cite{Susstrunk-Huber-2015}, but our goal is to design an interface to find EGV points.

Since the two sub-lattices have opposite $x$--$y$ cross-couplings, we require the interface connecting the two sub-lattices to have no $x$--$y$ cross-couplings to maintain neutrality. For each row, since the two sub-lattices have equal $x$--$x$ and $y$--$y$ couplings, we require the interface to have equal $x$--$x$ and $y$--$y$ couplings with the same sign. Thus, we denote the interface couplings as $\epsilon_S^{(0)}$ for the row $(x^{(0)}_{r,S},y^{(0)}_{r,S})$, $-\epsilon_S^{(1)}$ for the row $(x^{(1)}_{r,S},y^{(1)}_{r,S})$, and $-\epsilon_S^{(2)}$ for the row $(x^{(2)}_{r,S},y^{(2)}_{r,S})$, with the coupling strengths $\epsilon_S^{(j)}\geq0$ for $j=0,1,2$; see Fig.~\ref{fig:lattice_interface}. The interface is defined by the three free parameters $\epsilon_S^{(j)}$, $j=0,1,2$, but practically we assume $\epsilon_S^{(1)}=\epsilon_S^{(2)}$ consistent with the two sub-lattices and tune the two free parameters $\epsilon_S^{(0)}$ and $\epsilon_S^{(1)}$.

Finally, we require the interface to have the same self-coupling coefficient $A_s=3+\sqrt{3}$ as the two sub-lattices. To reach the total self-coupling $-A_sf$ possibly using springs attached to walls, the interface parameters must satisfy $\epsilon_S^{(0)} \in [0,\sqrt{3} f]$ and $\epsilon_S^{(1,2)} \in [0, (1+\sqrt{3})f/2]$. We will explore this compact parameter space to calculate the dispersion relation of the MTI and find EGV points.

The two sub-lattices are decoupled when $\epsilon_S^{(j)}=0$ for all $j$. In this limit, the dispersion relation $\alpha(k)$ of the MTI is simply two copies of Fig.~\ref{fig:dispersion_relation}. At each point $(k,\alpha)$ in a band gap, the four eigenvectors are the left and right edge states for the two sub-lattices, i.e., two edge states near the interface and two edge states near the vacuum. For small $\epsilon_S^{(j)}$, matrix perturbation theory reveals that the two edge states near the interface recombine into two interface states, one odd and one even, and the corresponding eigenvalues are distinct and differ from $\alpha$. We shall not show this calculation explicitly since EGV points yielding favorable CNLS parameters typically require $\epsilon_S^{(j)}=O(1)$ and can only be found numerically.

\begin{figure*}
\centering
\includegraphics[width=.75\textwidth]{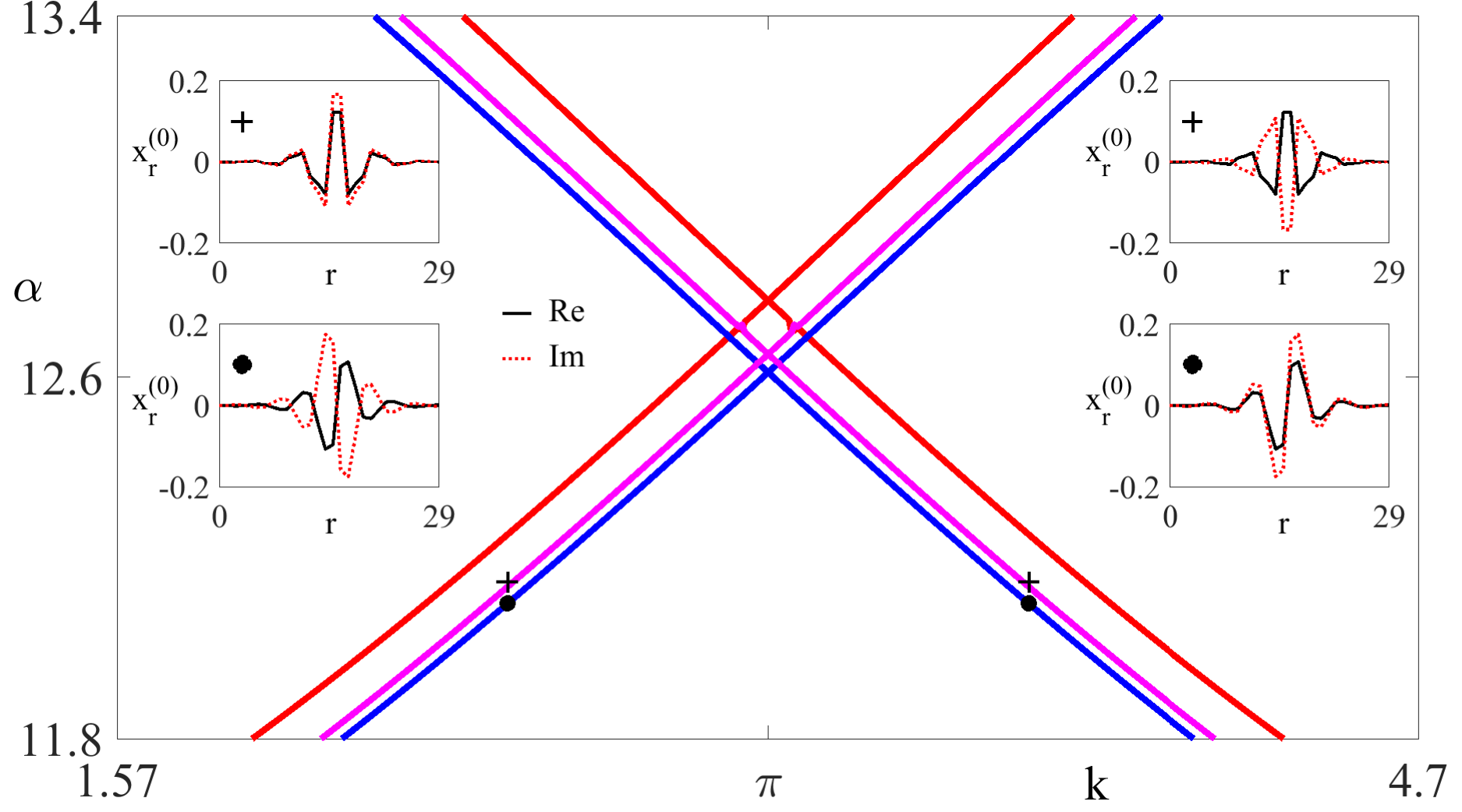}
    \caption{(Color online) The dispersion relation in the lower band gap with crossing point at $k=\pi$ as in Fig.~\ref{fig:dispersion_relation} but with interface couplings included. The dispersion relation of the two edge states near the vacuum, i.e., on the left edge of the left sub-lattice and on the right edge of the right sub-lattice, is shown in red. The dispersion relation of the interface states is shown in blue for the odd states and magenta for the even states. The two interface states at a generic $k$ and their Kramers partners at $2\pi-k$ are marked by `+' and `o', and the $x^{(0)}$ components of the corresponding eigenvectors are plotted in the sub-panels.}
\label{fig:DR_adapt}
\end{figure*}

The dispersion relation in the lower band gap is shown in Fig.~\ref{fig:DR_adapt} where $\epsilon_S^{(0)}=\Omega_0 f$ and $\epsilon_S^{(1)}=\epsilon_S^{(2)}=\Omega_0 f/2$ with $\Omega_0=1.5$. The dispersion relation $\alpha(k)$ of the two edge states near the vacuum is unchanged from Fig.~\ref{fig:dispersion_relation} as shown in red. The dispersion relation of the two interface states originates from splitting the two-fold degeneracy in $\alpha(k)$ and consists of the blue and magenta curves. As shown in the sub-panels, the blue curves contain the odd states, while the magenta curves contain the even states. We also see that the interface states at a generic point $(k,\alpha)$ and its Kramers partner $(2\pi-k,\alpha)$ are related by complex conjugation. Note that this parameter choice does not yield an EGV point but shows the general shape of the dispersion relation. Next, we will find parameter choices yielding EGV points, but we will not show the dispersion relations explicitly since they resemble Fig.~\ref{fig:DR_adapt}.

\section{Vector edge solitons and domain walls}\label{sec:ves-dw}
Hereafter, EGV points are always found between the odd and even interface states in the same band gap. Since both carrier frequencies $\alpha_0,\beta_0>0$ and the pendulum nonlinearity $\sigma>0$, the nonlinear coefficients of Eq.~(\ref{eq:MTI_CNLS_EQN1}) are always positive. Thus, Eq.~(\ref{eq:MTI_CNLS_EQN1}) is focusing if the dispersion coefficients are positive, i.e., $\alpha_0'',\beta_0''>0$, and defocusing if the dispersion coefficients are negative, i.e., $\alpha_0'',\beta_0''<0$. To find a vector edge soliton or DW, we solve numerically the original 2D MTI system (\ref{eq:eqn-of-motion}) with the following initial condition obtained from Eq.~(\ref{eq:CM_ansatz}):
\begin{equation}\label{eq:general-initial-condition}
\boldsymbol{X}_{r,S}(0)=\epsilon\left[ \mathcal{A}(\tilde{S})\boldsymbol{X}_r^{(1)}+ \mathcal{B}(\tilde{S})\boldsymbol{X}_r^{(2)} \right]e^{iSk_0}+c.c.,
\end{equation}
where $\mathcal{A}(\tilde{S})$ and $\mathcal{B}(\tilde{S})$ yield a vector soliton or DW in Eq.~(\ref{eq:MTI_CNLS_EQN1}), and $\boldsymbol{X}_r^{(1,2)}$ are the two interface states. Hereafter, we fix $\epsilon=0.1$ unless otherwise stated. We choose ${\cal A}(\tilde{S})$ and ${\cal B}(\tilde{S})$ to yield stable propagation of the profile over a long time ($\tau>100$) in Eq.~\eqref{eq:MTI_CNLS_EQN1}, but we do not insist that the profile remains stable for $\tau\to\infty$.

The rectangular lattice has $N_r \times N_s$ sites, i.e., $N_r\times N_S$ cells where $N_S=N_s/3$. The index $r$, $s$, or $S$ begins with 0. The interface is between $r=N_r/2-1$ and $N_r/2$, so the two sub-lattices have equal sizes. We assume periodicity in the $s$-direction and vacuum for $r\notin[0,N_r-1]$.

\subsection{Focusing CNLS coefficients}\label{sec:focu-cnls}
The interface couplings $\epsilon_S^{(0)}=\Omega_0 f$ and $\epsilon_S^{(1)}=\epsilon_S^{(2)}=\Omega_0 f/2$ with $\Omega_0=0.994$ yields an EGV point in the lower band gap at the carrier wavenumber $k_0=2.011$. The carrier frequencies are $(\alpha_0,\beta_0)=(11.797,11.805)$, and the group velocity is $\alpha_0'=\beta_0'=0.697$. This EGV point yields focusing CNLS coefficients for Eq.~(\ref{eq:MTI_CNLS_EQN1}) with
\begin{align*}
    &\alpha''_0=0.2096, \ \tilde{\sigma}_1=0.0072, \ \tilde{\sigma}_2=0.0061, \\
    &\beta''_0=0.2107,\ \tilde{\sigma}_3=0.0064, \ \tilde{\sigma}_4=0.0061.
\end{align*}

This CNLS equation is nearly symmetric, but we follow a general program to find localized solutions \cite{snee2024domain}. A key solution in the focusing regime is the scalar BB soliton, i.e., a sech function times a constant vector. In the 3D parameter space of the group velocity and the frequencies of the two components, such solitons form a codimension-1 family defined by a consistency condition. This family can be numerically continued into a codimension-0 family of BB solitons whose profiles are not a sech function, but we do not show them explicitly.

We choose the two free parameters as the group velocity $C_g$ and the shifted frequency $\gamma$ of the ${\cal A}$ component. We also include two free parameters $\Theta_{\mathcal{A},\mathcal{B}}$ for the phases (polarizations) of the two bright components. Although these phases do not affect propagation of a single soliton, they are crucial for collisions between two solitons. Thus, scalar BB solitons can be written as
\begin{equation}\label{eq:BB_ansatz_A}
\begin{aligned}
 &\mathcal{A}(\tilde{S},\tau)=\Lambda_\mathcal{A} W \mbox{sech}(W [ \tilde{S}-C_g\tau])\mbox{e}^{i\Phi_\mathcal{A}}, \\
 &\mathcal{B}(\tilde{S},\tau)=\Lambda_\mathcal{B} W \mbox{sech}(W[ \tilde{S}-C_g\tau])\mbox{e}^{i\Phi_\mathcal{B}},
\end{aligned}
\end{equation}
where
\begin{equation}\label{eq:BB-Phi}
\Phi_\mathcal{A}=\frac{C_g}{\alpha_0''}\tilde{S}-\nu_\mathcal{A}\tau+\Theta_\mathcal{A},\quad
\Phi_\mathcal{B}=\frac{C_g}{\beta_0''}\tilde{S}-\nu_\mathcal{B}\tau+\Theta_\mathcal{B}.
\end{equation}
Here, the consistency condition relates the frequency of the ${\cal B}$ component to the frequency of the ${\cal A}$ component:
\begin{equation}\label{eq:BB-nu}
\nu_\mathcal{A}=\frac{C_g^2}{\alpha_0''}+\gamma, \quad
\nu_\mathcal{B}=\frac{C_g^2}{2}\left(\frac{\beta_0''^2+\alpha_0''^2}{\alpha_0''^2\beta_0''}\right)+\frac{\beta_0''}{\alpha_0''}\gamma,
\end{equation}
the scaling factors depend only on the CNLS coefficients:
\begin{gather*}
 \Lambda_\mathcal{A}=\sqrt{\frac{\tilde{\sigma}_3\alpha_0''-2\tilde{\sigma}_2\beta_0''}{3\tilde{\sigma}_1\tilde{\sigma}_3-12\tilde{\sigma}_2\tilde{\sigma}_4}}, \quad
 \Lambda_\mathcal{B}=\sqrt{\frac{\tilde{\sigma}_1\beta_0''-2\tilde{\sigma}_4\alpha_0''}{3\tilde{\sigma}_1\tilde{\sigma}_3-12\tilde{\sigma}_2\tilde{\sigma}_4}},
\end{gather*}
and $(C_g,\gamma)$ must be chosen to make the amplitude real:
\begin{equation*}
     W=\sqrt{-\left(\frac{C_g^2}{\alpha_0''^2}+\frac{2\gamma}{\alpha_0''}\right)}.
\end{equation*}

\begin{figure}
\centering
\subfigure[]{\includegraphics[width=.23\textwidth]{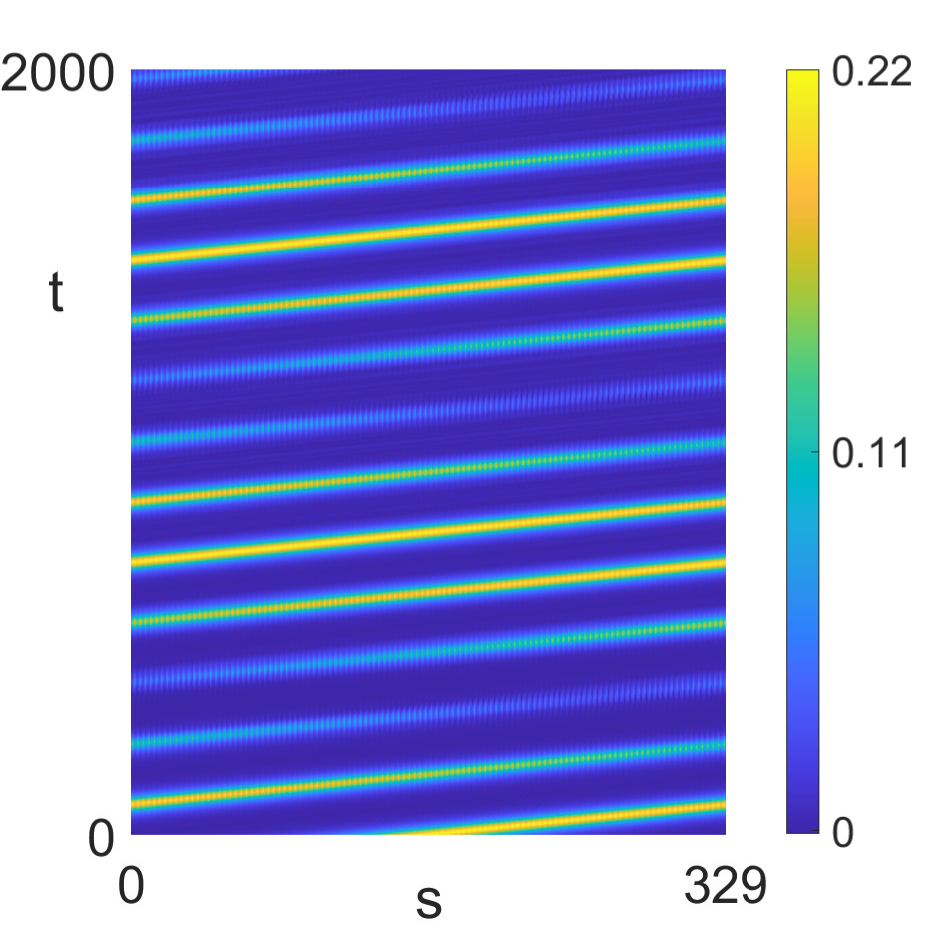}\label{fig:MTI2_BB_Left}}
\subfigure[]{\includegraphics[width=.23\textwidth]{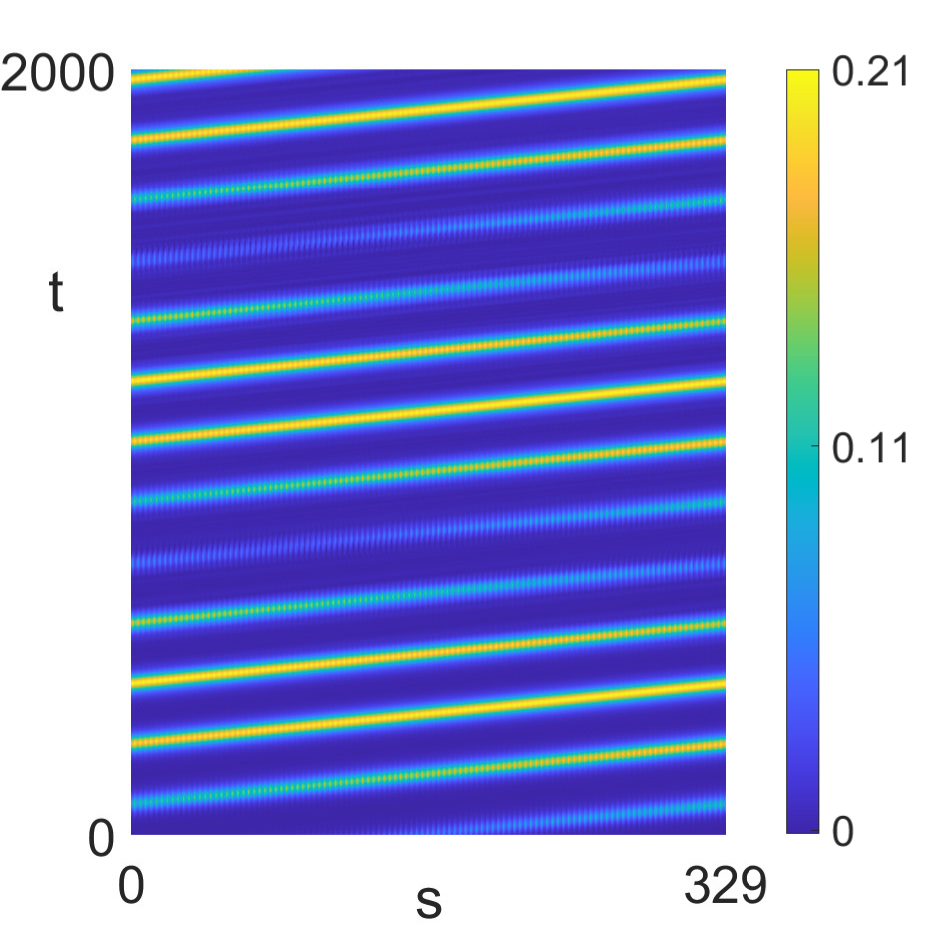}\label{fig:MTI2_BB_Right}}\\
\subfigure[]{\includegraphics[width=.18\textwidth]{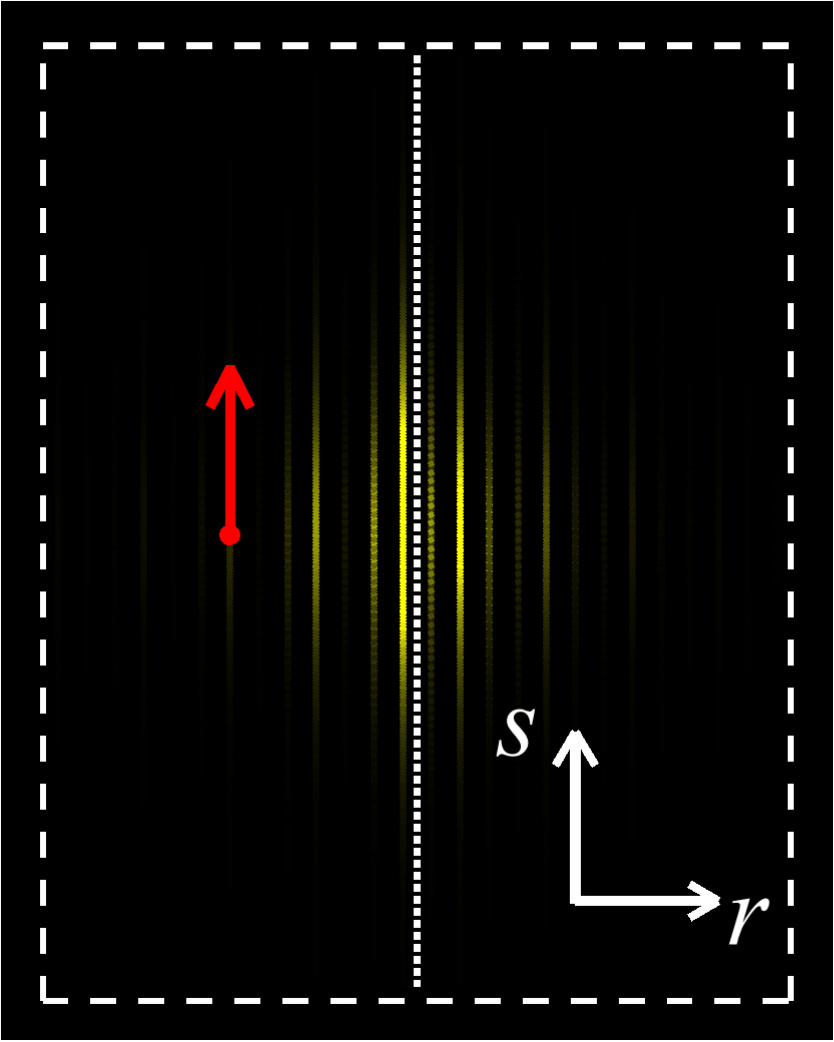}\label{fig:MTI2_BB_2Ddomain}}
\subfigure[]{\includegraphics[width=.23\textwidth]{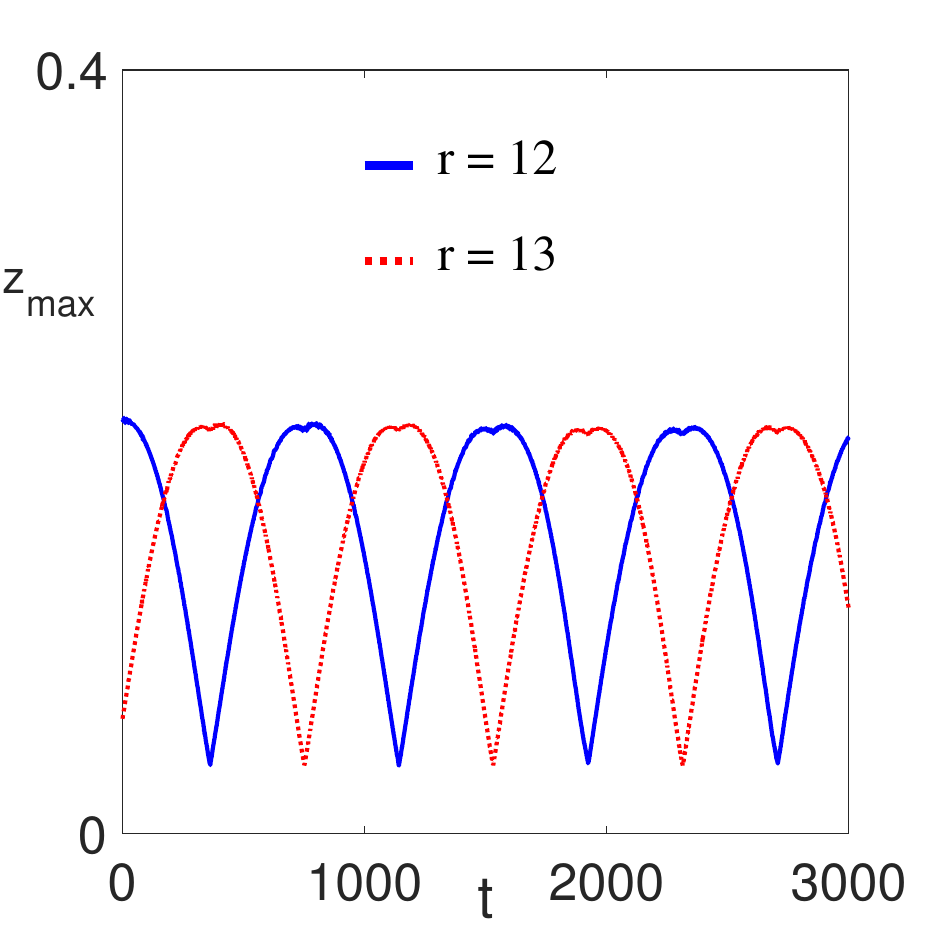}\label{fig:MTI2_BB_MAXAMP}}
    \caption{(Color online) BB edge soliton propagating along the interface of the MTI with two topological sectors. The lattice size is $N_r \times N_s=26 \times 330$. (a,b) Space-time plots of the interface sites at $r=12$ and $r=13$ respectively. (c) 2D snapshot showing the initial condition for the BB edge soliton. White-dashed lines are the lattice boundaries, white-dotted line is the interface between the two sub-lattices, and the red arrow indicates the propagation direction. (d) Maximum amplitude vs time for the $r=12$ and $r=13$ sites.}
    \label{fig:BB_solution}
\end{figure}

We set $\tau=0$ in Eq.~(\ref{eq:BB_ansatz_A}), substitute into Eq.~(\ref{eq:general-initial-condition}), and solve the governing equation (\ref{eq:eqn-of-motion}) of the MTI numerically on a domain of $N_r \times N_s= 26 \times 330$ sites periodic in the $s$-direction to obtain a BB edge soliton propagating along the interface. Figure \ref{fig:BB_solution} shows the solution over a long time using the parameters $(C_g,\gamma,\Theta_\mathcal{A},\Theta_\mathcal{B})=(0,-0.15,0,0)$. Any 2D spatial profile constructed using Eq.~\eqref{eq:general-initial-condition} decays into the bulk of the two sub-lattices; see e.g.~the initial condition for the BB edge soliton in Fig.~\ref{fig:MTI2_BB_2Ddomain}. Thus, we can effectively show the dynamics of the MTI using space-time plots of the left and right interface sites. For the BB edge soliton, such plots in Fig.~\ref{fig:MTI2_BB_Left} and Fig.~\ref{fig:MTI2_BB_Right} show that this nonlinear wave packet propagates stably due to the balance between dispersion and nonlinearity. Note that in all such plots, the variable shown at each site is the amplitude $z=\sqrt{x^2+y^2}$, not $x$ or $y$ individually.

An emergent property of the 2D MTI dynamics absent in the 1D CNLS dynamics is a time-periodic beat due to the linear superposition of the two interface states with different carrier frequencies $\alpha_0$ and $\beta_0$; see the multiple scale ansatz (\ref{eq:CM_ansatz}). For the BB edge soliton, the period of this beat is $2\pi/(\alpha_0-\beta_0) \approx 778.9$, which agrees with the plot of the maximum amplitude over time in Fig.~\ref{fig:MTI2_BB_MAXAMP}. The set of $\tilde{S}$ exhibiting this beat satisfies $\mathcal{A}(\tilde{S})\mathcal{B}(\tilde{S})\neq0$. For BB edge solitons, this set is the center of the sech envelope. Thus, a BB edge soliton resembles a bright breather, i.e., a localized oscillation embedded in a zero background. Note that a similar beat occurs in vector edge solitons in photonic or cold atomic TIs \cite{ivanov2020vector,ivanov2021topological,tian2023vector}.

\subsection{Defocusing CNLS coefficients}\label{sec:defo-cnls}
The interface couplings $\epsilon_S^{(0)}=\epsilon_S^{(1)}=\epsilon_S^{(2)}=\Omega_0 f/2$ with $\Omega_0=0.698$ yield an EGV point in the upper band gap at the carrier wavenumber $k_0=\pi/10$. The carrier frequencies are $(\alpha_0,\beta_0)=(16.537,16.766)$, and the group velocity is $\alpha'_0=\beta'_0=0.617$. This EGV point yields defocusing CNLS coefficients for Eq.~(\ref{eq:MTI_CNLS_EQN1}) with
\begin{align*}
    &\alpha''_0=-0.1335, \ \tilde{\sigma}_1=0.0062, \ \tilde{\sigma}_2=0.0060, \\
    &\beta''_0=-0.1428,\ \tilde{\sigma}_3=0.0061, \ \tilde{\sigma}_4=0.0059.
\end{align*}

This CNLS equation is nearly symmetric, but we again follow a general program to find localized solutions \cite{snee2024domain}. In the BEC setting, vector solitons in the defocusing CNLS equation are extensively studied \cite{kevrekidis2016solitons}. Most such studies assume that the two dispersion coefficients are equal, while the four nonlinear coefficients are arbitrary.

We first seek DD solitons as the defocusing counterpart of BB solitons in the focusing regime. In BECs, the formation of DD solitons can be predicted from the interaction of dark solitons in different components \cite{ohberg2001dark}. Here, we seek the scalar DD soliton, i.e., a tanh function times a constant vector. As before, such solitons form a codimension-1 family defined by a consistency condition, and we choose the two free parameters as the group velocity $C_g$ and the shifted frequency $\gamma$ of the ${\cal A}$ component. Thus, scalar DD solitons can be written as
\begin{equation}\label{eq:tanh_1}
\begin{aligned}
 &\mathcal{A}(\tilde{S},\tau)=\Gamma_\mathcal{A} \mathcal{W} \mbox{tanh}(\mathcal{W} [ \tilde{S}-C_g\tau])\mbox{e}^{i\Phi_\mathcal{A}}, \\
 &\mathcal{B}(\tilde{S},\tau)=\Gamma_\mathcal{B} \mathcal{W} \mbox{tanh}(\mathcal{W}[ \tilde{S}-C_g\tau])\mbox{e}^{i\Phi_\mathcal{B}}.
\end{aligned}
\end{equation}
Here, $\Phi_{\mathcal{A},\mathcal{B}}$ are given by Eq.~\eqref{eq:BB-Phi} but with $\Theta_{\mathcal{A},\mathcal{B}}=0$ since a dark component has no phase freedom during soliton collisions. The consistency condition still relates the frequency of the ${\cal B}$ component to the frequency of the ${\cal A}$ component via Eq.~\eqref{eq:BB-nu}. The scaling factors depend only on the CNLS coefficients:
\begin{equation*}
 \Gamma_\mathcal{A}=\sqrt{\frac{2\tilde{\sigma}_2\beta_0''-\tilde{\sigma}_3\alpha_0''}{3\tilde{\sigma}_1\tilde{\sigma}_3-12\tilde{\sigma}_2\tilde{\sigma}_4}}, \quad
 \Gamma_\mathcal{B}=\sqrt{\frac{2\tilde{\sigma}_4\alpha_0''-\tilde{\sigma}_1\beta_0''}{3\tilde{\sigma}_1\tilde{\sigma}_3-12\tilde{\sigma}_2\tilde{\sigma}_4}},
\end{equation*}
and $(C_g,\gamma)$ must be chosen to make the amplitude real:
\begin{equation*}
 \mathcal{W}=\sqrt{\frac{C_g^2}{2\alpha_0''^2}+\frac{\gamma}{\alpha_0''}}.
\end{equation*}
Note that the two-parameter family of scalar DD solitons in Eq.~\eqref{eq:tanh_1} can be numerically continued in the frequency of the ${\cal B}$ component and the wavenumbers of the two components into a five-parameter family of DD solitons, but we will not show them explicitly.

The 2D MTI dynamics exhibit a time-periodic beat with period $T_0\equiv2\pi/(\alpha_0-\beta_0) \approx 27.4$. To remove this beat and thus reconstruct the 1D CNLS dynamics from the 2D MTI dynamics, we can use the parity of the two interface states. If $\boldsymbol{X}_r^{(1)}$ and $\boldsymbol{X}_r^{(2)}$ are the even and odd interface states respectively, then $\boldsymbol{X}_{\frac{N_r}{2}}^{(1)}=\boldsymbol{X}_{\frac{N_r}{2}-1}^{(1)}$ and $\boldsymbol{X}_{\frac{N_r}{2}}^{(2)}=-\boldsymbol{X}_{\frac{N_r}{2}-1}^{(2)}$ at the interface. To separate the contributions of ${\cal A}$ and ${\cal B}$, we recombine each pair of interface pendula: $x_{\mathcal{A},\mathcal{B}}=(x_{\frac{N_r}{2}}\pm x_{\frac{N_r}{2}-1})/4$ and $y_{\mathcal{A},\mathcal{B}}=(y_{\frac{N_r}{2}}\pm y_{\frac{N_r}{2}-1})/4$. Then, we combine $x$ and $y$ into a single amplitude to yield two excitation variables for each pair of interface sites: $z_{\mathcal{A},\mathcal{B}}=\sqrt{x_{\mathcal{A},\mathcal{B}}^2+y_{\mathcal{A},\mathcal{B}}^2}$, such that $(z_{{\cal A},{\cal B}})_s$ are proportional to $|{\cal A}(\tilde{S})|$ and $|{\cal B}(\tilde{S})|$.

\begin{figure}
\centering
\subfigure[]{\includegraphics[width=.23\textwidth]{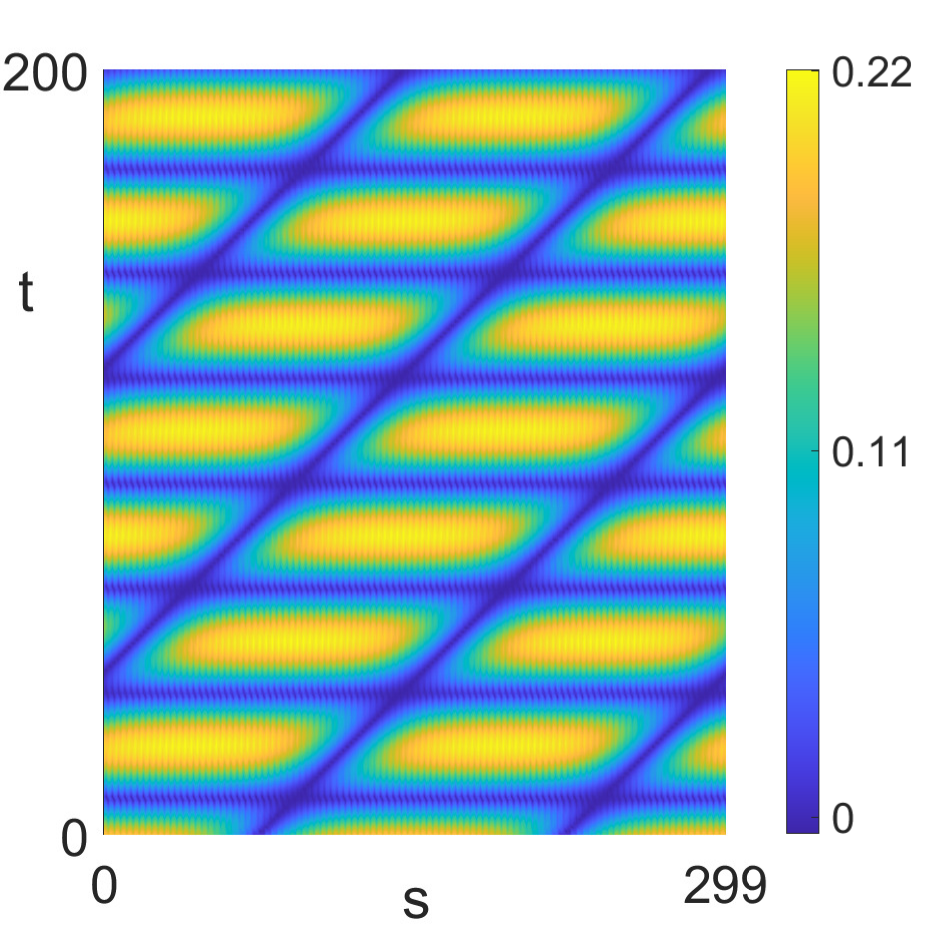}\label{fig:MTI2_DD_Left}}
\subfigure[]{\includegraphics[width=.23\textwidth]{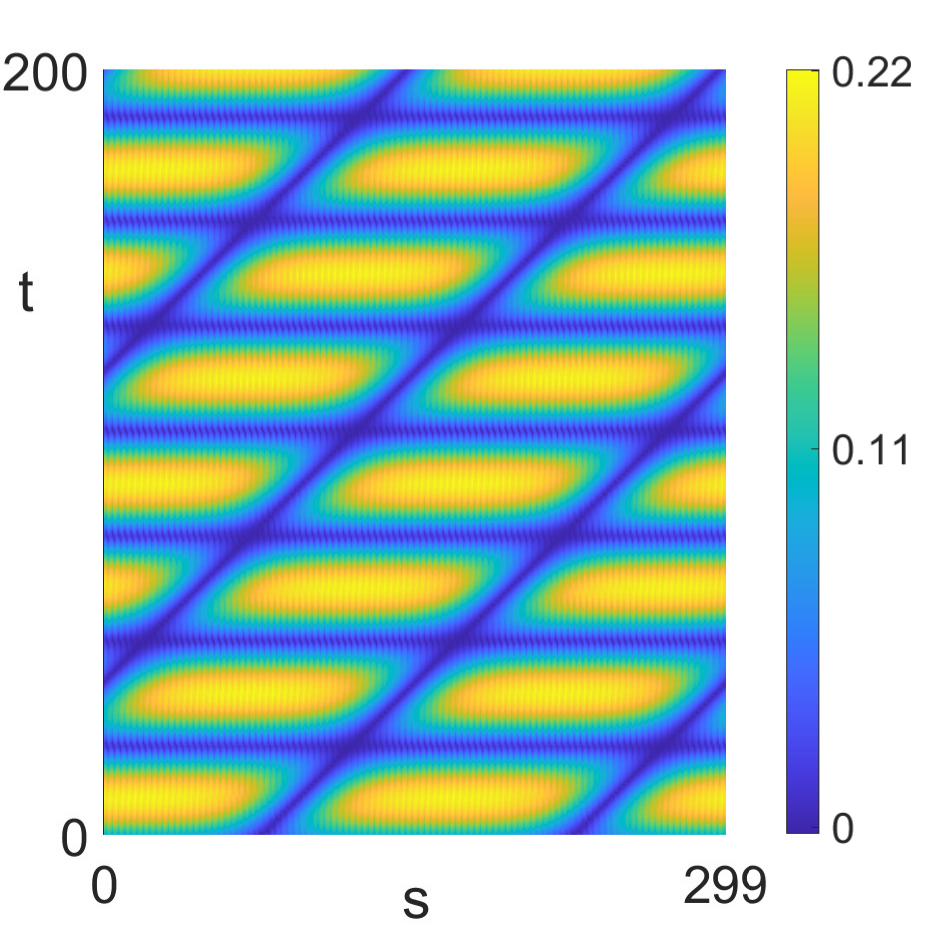}\label{fig:MTI2_DD_Right}}\\
\subfigure[]{\includegraphics[width=.23\textwidth]{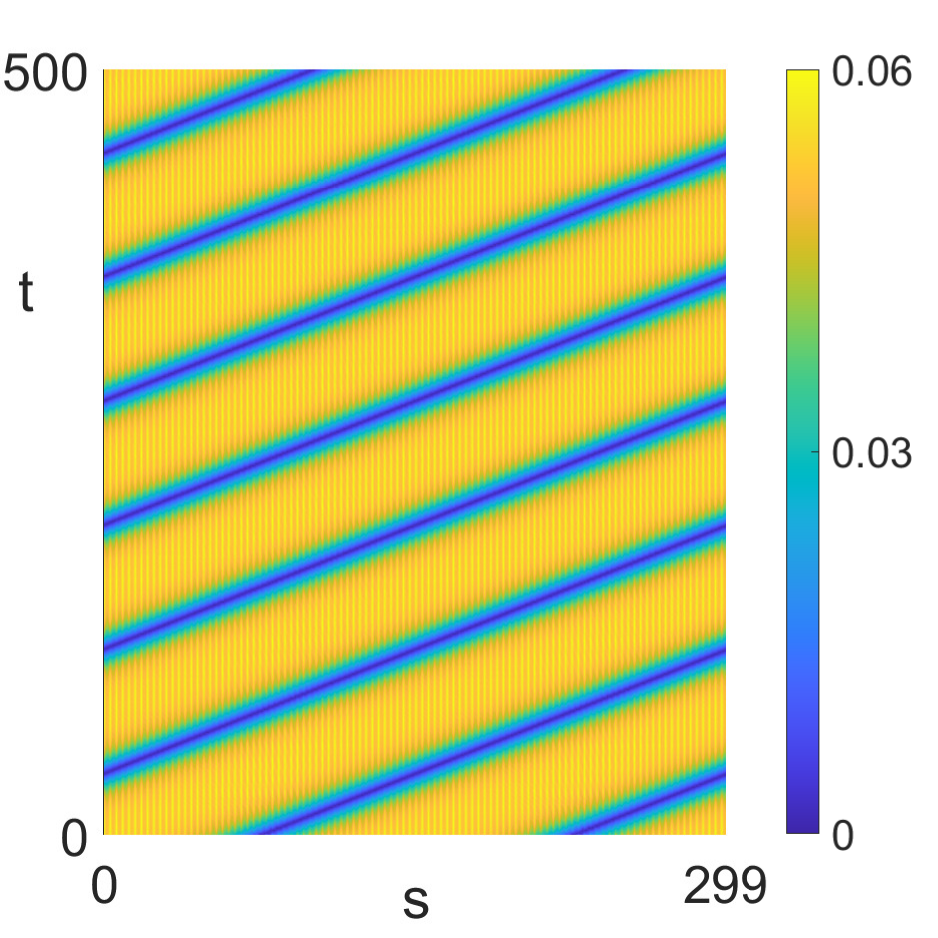}\label{fig:MTI2_DD_zA}}
\subfigure[]{\includegraphics[width=.23\textwidth]{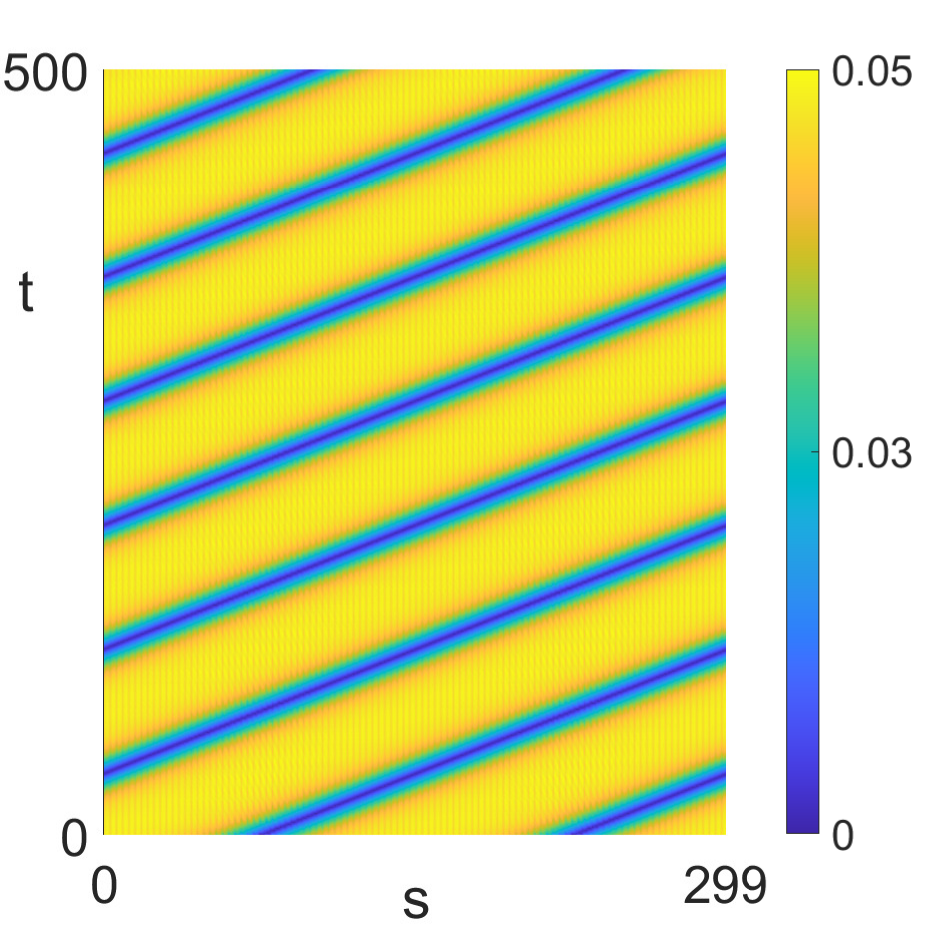}\label{fig:MTI2_DD_zB}}\\
\subfigure[]{\includegraphics[width=.23\textwidth]{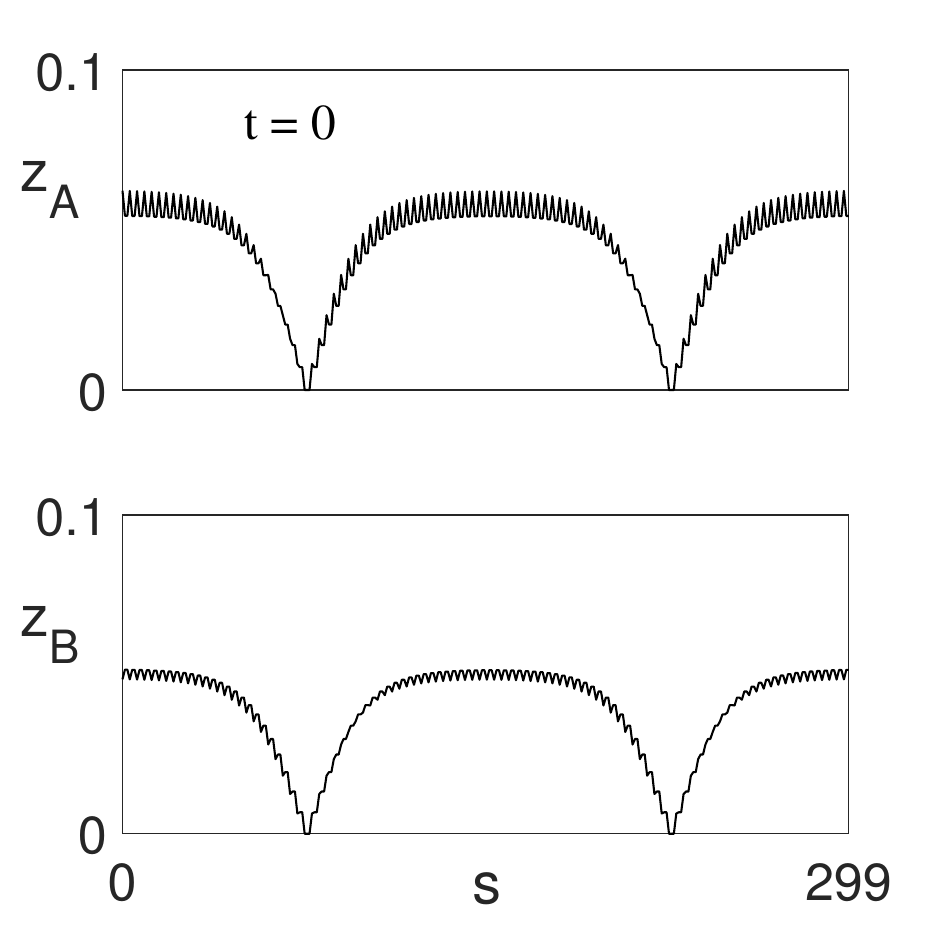}\label{fig:MTI2_DD_init}}
\subfigure[]{\includegraphics[width=.23\textwidth]{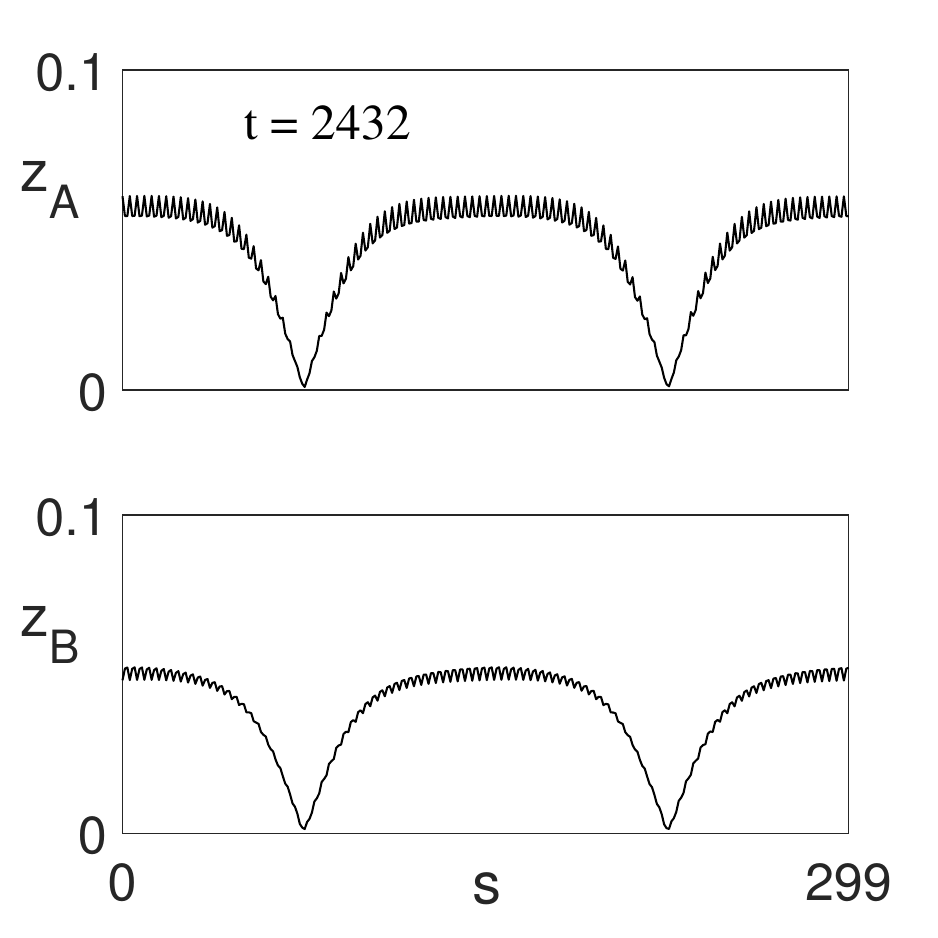}\label{fig:MTI2_DD_longtime}}
    \caption{(Color online) Two DD edge solitons propagating along the interface of the MTI with $N_r \times N_s = 26 \times 300$ sites. (a,b) Space-time plots of the interface sites at $r=12$ and $r=13$ respectively. (c,d) Space-time plots of the excitation variables $z_\mathcal{A}$ and $z_\mathcal{B}$. (e,f) Profiles of the excitation variables $z_\mathcal{A}$ and $z_\mathcal{B}$ initially and after a long time.}
    \label{fig:DD_solution}
\end{figure}

For localized solutions with nonzero backgrounds, we choose the lattice size as $N_r \times N_s = 26 \times 300$, such that the number of cells $N_S=N_s/3=100$ is an integer multiple of the carrier wavelength $2\pi/k_0=20$. Since the background of a scalar DD soliton $({\cal A}(\tilde{S}),{\cal B}(\tilde{S}))$ in Eq.~(\ref{eq:tanh_1}) changes sign as $\tilde{S}\to\pm\infty$, we pair this soliton with its partner $(-{\cal A}(\tilde{S}),-{\cal B}(\tilde{S}))$ to maintain periodicity in $\tilde{S}$. Figure \ref{fig:DD_solution} shows two DD edge solitons propagating along the interface of the MTI using the parameters $(C_g,\gamma)=(0,-0.23)$. As shown in Figs.~\ref{fig:MTI2_DD_Left} and \ref{fig:MTI2_DD_Right}, in terms of the amplitude $z$ of the interface sites, a DD edge soliton exhibits a beat with period $T_0$ and resembles a dark breather. Since a DD soliton in 1D CNLS has nonzero backgrounds in both ${\cal A}$ and ${\cal B}$, the associated dark breather in 2D MTI has an oscillating background.

The DD soliton in Eq.~\eqref{eq:tanh_1} has zero centers in both ${\cal A}$ and ${\cal B}$ and is also called a black-black soliton. Thus, the associated dark breather in Figs.~\ref{fig:MTI2_DD_Left} and \ref{fig:MTI2_DD_Right} is a black breather with a zero center. The black-black soliton in Eq.~\eqref{eq:tanh_1} can be continued in the wavenumber(s) of ${\cal A}$ and/or ${\cal B}$ into black-gray or gray-gray solitons, where gray means a nonzero center \cite{snee2024domain}. The dark breather associated with a black-gray soliton is a gray breather with a stationary nonzero center, while the dark breather associated with a gray-gray soliton is a gray breather with an oscillating center; these are not shown explicitly.

Despite this rich behavior, a DD edge soliton in terms of the excitation variables $z_\mathcal{A}$ and $z_\mathcal{B}$ simply exhibits the 1D CNLS dynamics as shown in Figs.~\ref{fig:MTI2_DD_zA} and \ref{fig:MTI2_DD_zB}. The reconstructed profile of $z_\mathcal{A}$ and $z_\mathcal{B}$ is almost unchanged over a long time as shown in Figs.~\ref{fig:MTI2_DD_init} and \ref{fig:MTI2_DD_longtime}, which shows the robust nature of such nonlinear waves. Note that the excitation variables $(z_{\mathcal{A},\mathcal{B}})_s$ in these panels keep the small variations between the three sites within a cell, which could be eliminated by an additional averaging.

Next, we seek DWs in our defocusing CNLS equation. In BECs, DWs require the immiscibility condition on the CNLS coefficients \cite{trippenbach2000structure,coen2001domain,alama2015domain}. For Eq.~\eqref{eq:MTI_CNLS_EQN1}, this condition is $4\tilde{\sigma}_2\tilde{\sigma}_4-\tilde{\sigma}_1\tilde{\sigma}_3>0$, which indeed holds.

To find DWs, we follow the steps in Appendix \ref{app:cnls}. The two fields $A$ and $B$ in Eq.~\eqref{Eq:GCNLS} correspond to ${\cal A}$ and ${\cal B}$ in Eq.~\eqref{eq:MTI_CNLS_EQN1}, so the coefficients $d_{1-2}$ and $g_{1-4}$ in Eq.~\eqref{Eq:GCNLS} can be identified with those in Eq.~\eqref{eq:MTI_CNLS_EQN1}. We can reduce Eq.~\eqref{Eq:GCNLS} to a real ordinary differential equation (ODE), Eq.~\eqref{eq:real-ode}, with two free parameters $\bomega_{\cal A}$ and $\bomega_{\cal B}$. In this ODE, a DW is a heteroclinic orbit between the two equilibria $ZN$ and $NZ$, where $Z$ ($N$) denotes a zero (nonzero) value. This orbit exists when $\bomega_{\cal B}/\bomega_{\cal A}=\sqrt{g_3g_4/(g_1g_2)}$ for Eq.~\eqref{Eq:GCNLS}, or equivalently for Eq.~\eqref{eq:MTI_CNLS_EQN1}:
\begin{equation}\label{eq:het_con}
\bomega_{\cal B}/\bomega_{\cal A}=\sqrt{\tilde{\sigma}_3\tilde{\sigma}_4/(\tilde{\sigma}_1\tilde{\sigma}_2)}.
\end{equation}
This condition and Eq.~\eqref{eq:bomega} define a 2D subset of the 3D parameter space of the group velocity $C_g$ and the frequencies of the two components $\omega_{\cal A}$ and $\omega_{\cal B}$. Here, we choose a point $(C_g,\omega_\mathcal{A},\omega_\mathcal{B})=(0,0.297,0.293)$ in this subset. A possible numerical method to find the DW at this point is to make Eq.~\eqref{eq:real-ode} the steady state of a dissipative partial differential equation and evolve an initial step between $ZN$ and $NZ$ to the steady state.

\begin{figure}
\centering
\subfigure[]{\includegraphics[width=.23\textwidth]{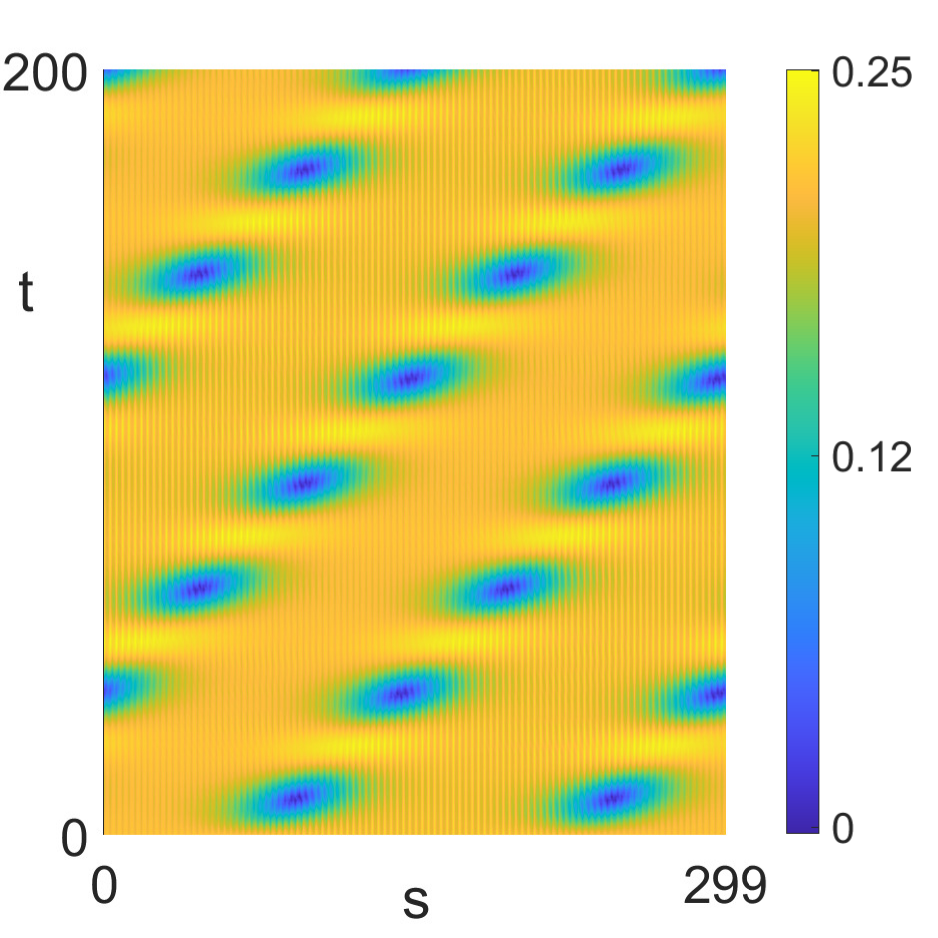}\label{fig:MTI2_front_Left}}
\subfigure[]{\includegraphics[width=.23\textwidth]{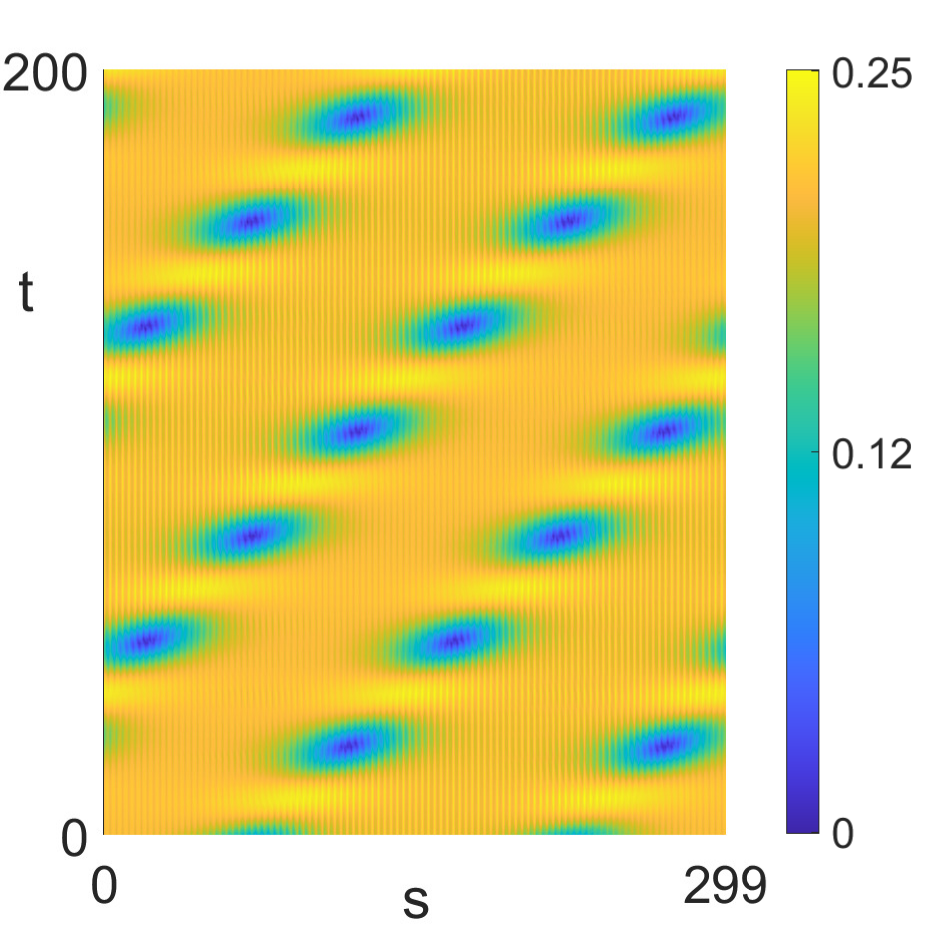}\label{fig:MTI2_front_Right}}\\
\subfigure[]{\includegraphics[width=.23\textwidth]{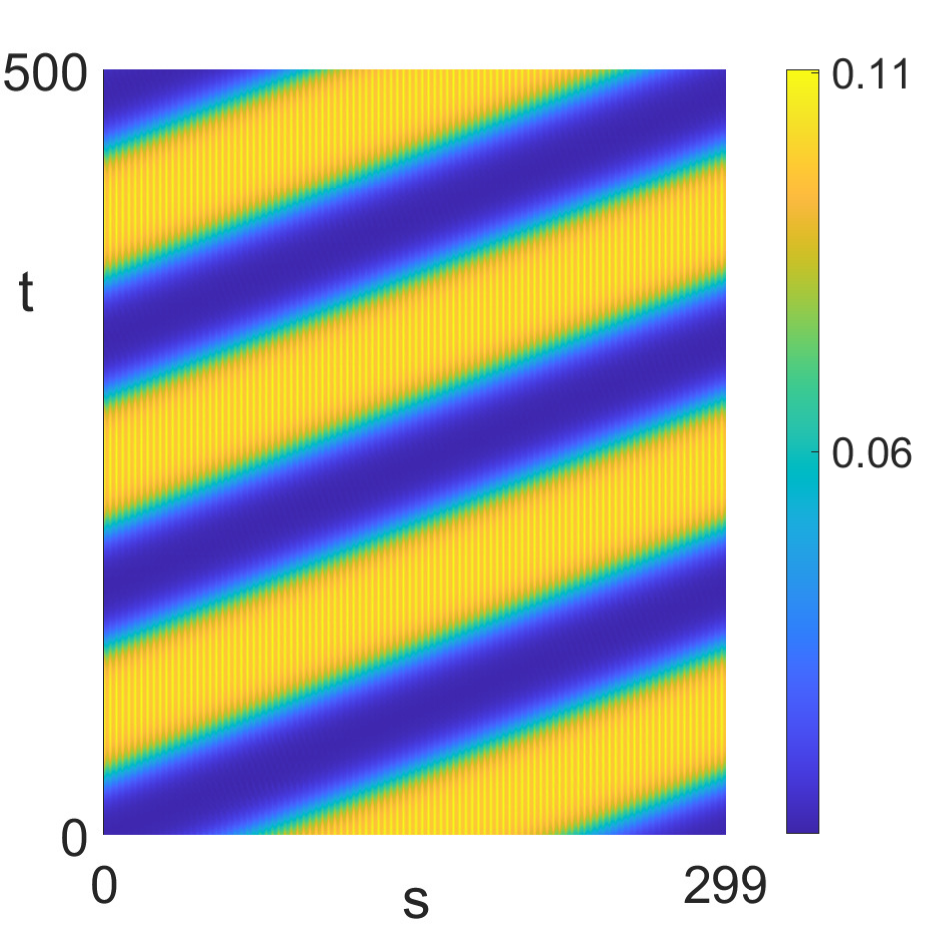}\label{fig:MTI2_front_zA}}
\subfigure[]{\includegraphics[width=.23\textwidth]{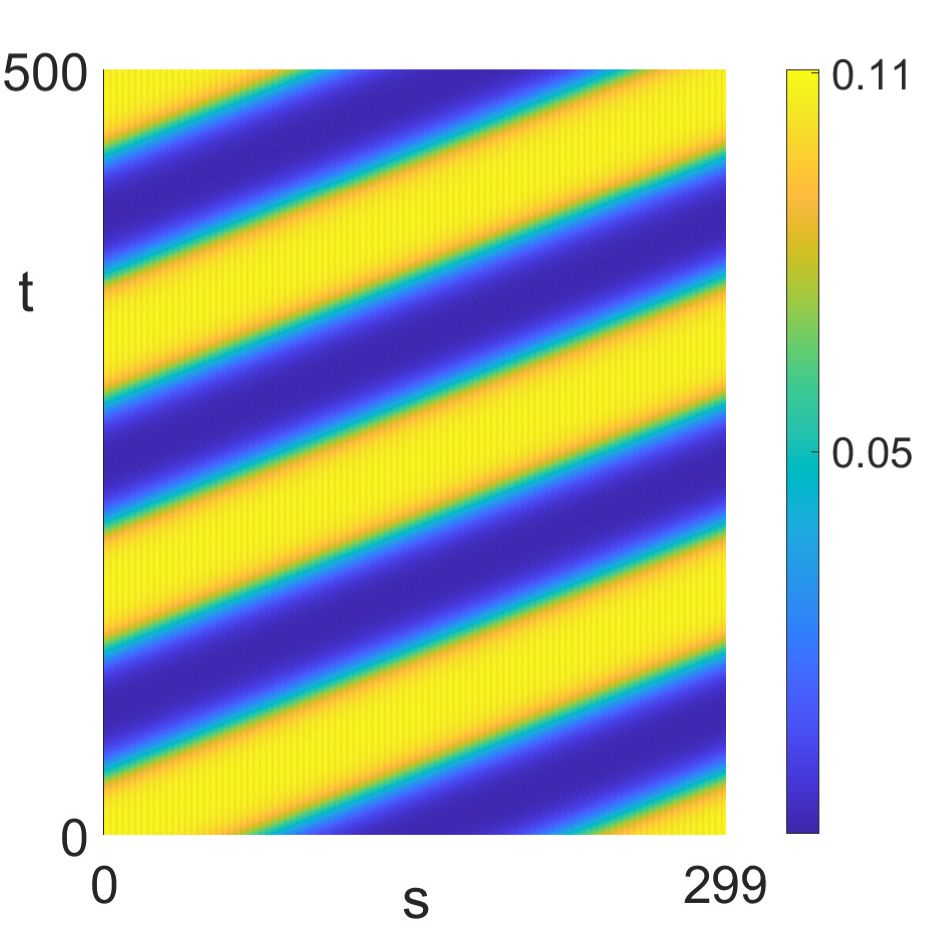}\label{fig:MTI2_front_zB}}\\
\subfigure[]{\includegraphics[width=.23\textwidth]{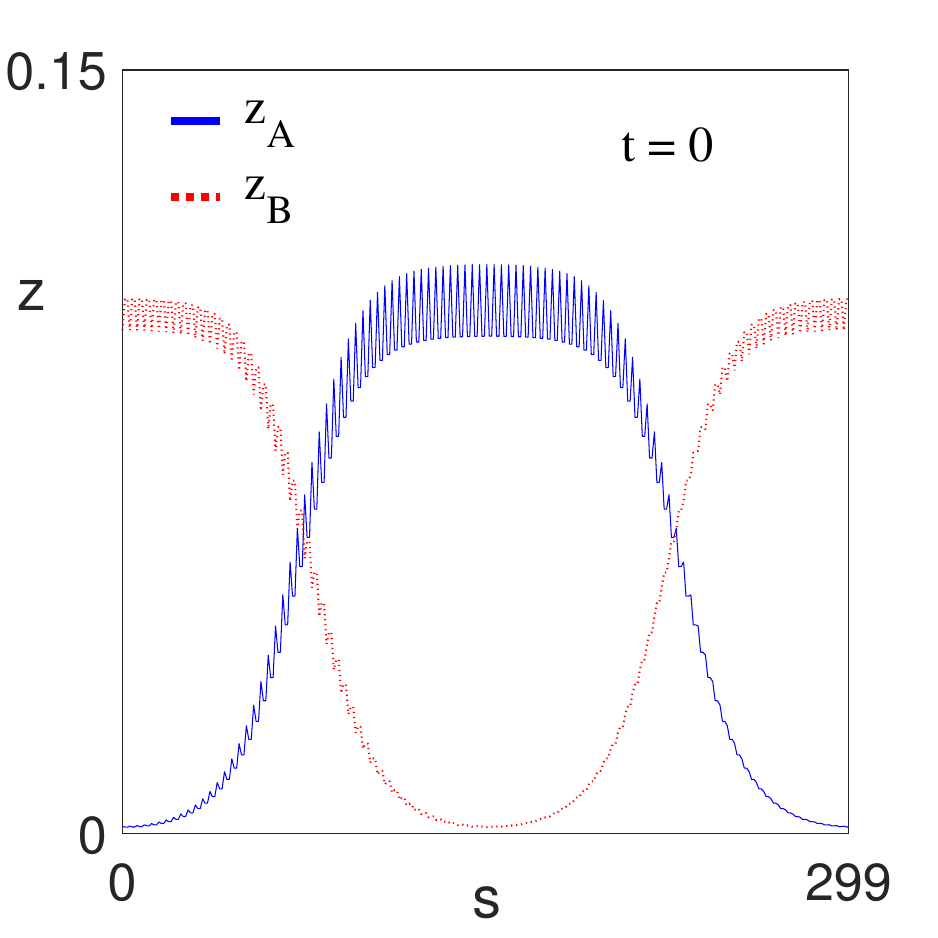}\label{fig:MTI2_front_init}}
\subfigure[]{\includegraphics[width=.23\textwidth]{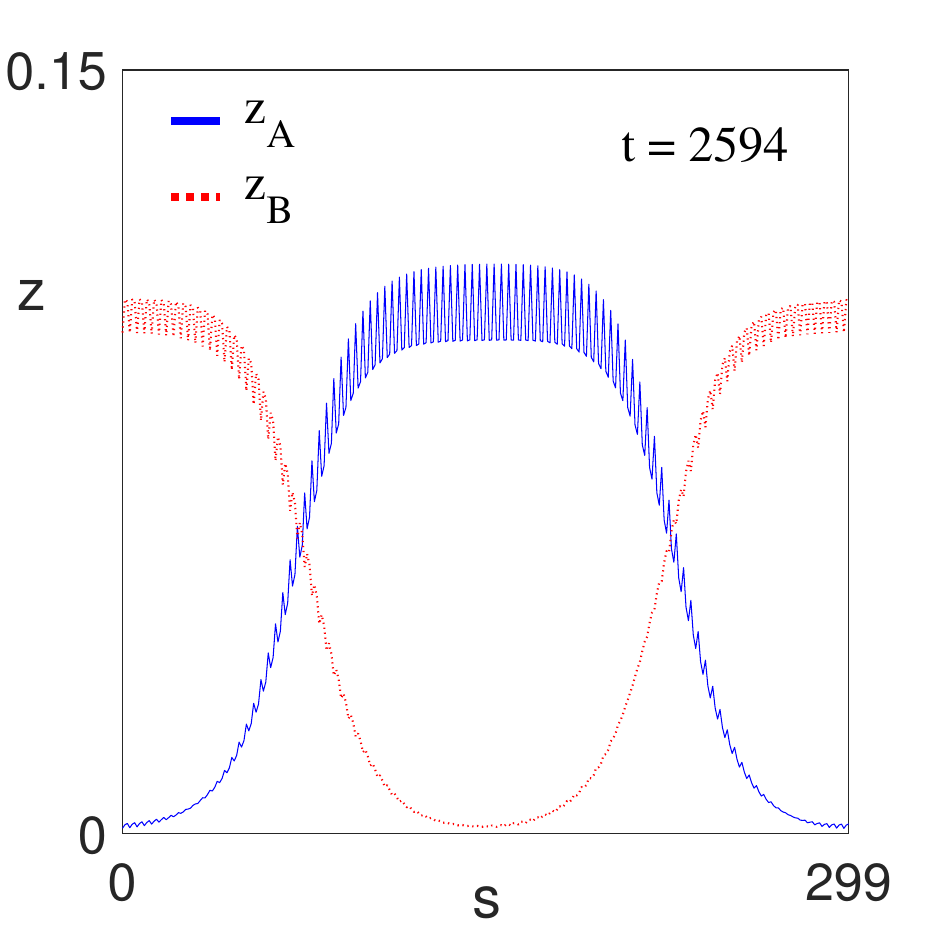}\label{fig:MTI2_front_longtime}}
    \caption{(Color online) Two edge DWs propagating along the interface of the MTI with $N_r \times N_s=26 \times 300$ sites. (a,b) Space-time plots of the interface sites at $r=12$ and $r=13$ respectively. (c,d) Space-time plots of the excitation variables $z_\mathcal{A}$ and $z_\mathcal{B}$. (e,f) Profiles of the excitation variables $z_\mathcal{A}$ and $z_\mathcal{B}$ initially and after a long time.}
    \label{fig:front_solution}
\end{figure}

To maintain periodicity in $\tilde{S}$, we pair a DW with its partner to form a heteroclinic cycle $ZN\to NZ\to ZN$. If $({\cal A}(\tilde{S}),{\cal B}(\tilde{S}))$ is a DW in 1D CNLS, then the solution to 2D MTI in Eq.~\eqref{eq:general-initial-condition} will be called an edge DW. Figure \ref{fig:front_solution} shows two edge DWs propagating along the interface of the MTI. As shown in Figs.~\ref{fig:MTI2_front_Left} and \ref{fig:MTI2_front_Right}, in terms of the amplitude $z$ of the interface sites, an edge DW connects two stationary nonzero backgrounds via a central beat with period $T_0$. These two backgrounds have almost equal amplitudes, but one consists of only the even state while the other consists of only the odd state. Thus, a single edge DW cannot be deemed a dark breather, but two edge DWs resemble a dark breather with a central plateau. An edge DW exhibits a transition in the parity of the background along the $s$-direction. Meanwhile, the MTI exhibits a transition between the two topological sectors along the $r$-direction. If we consider the parity of the background as a topological invariant, then an edge DW divides the MTI into four quarters with topological transitions along both $r$ and $s$ directions. As shown in Figs.~\ref{fig:MTI2_front_zA} and \ref{fig:MTI2_front_zB}, in terms of the excitation variables $z_\mathcal{A}$ and $z_\mathcal{B}$, an edge DW simply exhibits the 1D CNLS dynamics. As shown in Figs.~\ref{fig:MTI2_front_init} and \ref{fig:MTI2_front_longtime}, the profile of $z_\mathcal{A}$ and $z_\mathcal{B}$ is almost unchanged after a long time, which shows the robust nature of this solution.

Finally, we seek DB solitons in our defocusing CNLS equation. DB solitons are extensively studied since they exhibit ``symbiosis'': the bright soliton only exists in the defocusing regime due to an effective potential created by the dark soliton \cite{kevrekidis2016solitons}. In BECs, a pioneering study reveals the dynamics of DB solitons in a harmonic trap \cite{busch2001dark}. Experimentally, DB solitons are first realized using a phase imprinting method \cite{becker2008oscillations}. Subsequently, it is shown that the counterflow of two superfluids can even generate trains of DB solitons \cite{hamner2011generation}.

Here, we seek DB solitons parametrically related to DWs. The heteroclinic cycle $ZN\to NZ\to ZN$ formed by assembling two DWs is an extended DB soliton. This codimension-1 solution satisfying Eq.~\eqref{eq:het_con} can be numerically continued in Eq.~\eqref{eq:real-ode} into a codimension-0 family of DB solitons embedded in either $ZN$ or $NZ$. For the DW found above, continuation along the positive $\omega_{\cal A}$ direction yields a DB soliton at $(C_g,\omega_\mathcal{A},\omega_\mathcal{B})=(0,0.332,0.293)$ embedded in $ZN$. Continuation along negative $\omega_{\cal A}$ yields DB solitons embedded in $NZ$, which are not shown explicitly. Note that this three-parameter family of DB solitons can be numerically continued in the wavenumber of the dark component, while the bright component can be multiplied by a constant phase factor for soliton collisions; see Ref.~\cite{snee2024domain} for a numerical example.

\begin{figure}
\centering
\subfigure[]{\includegraphics[width=.23\textwidth]{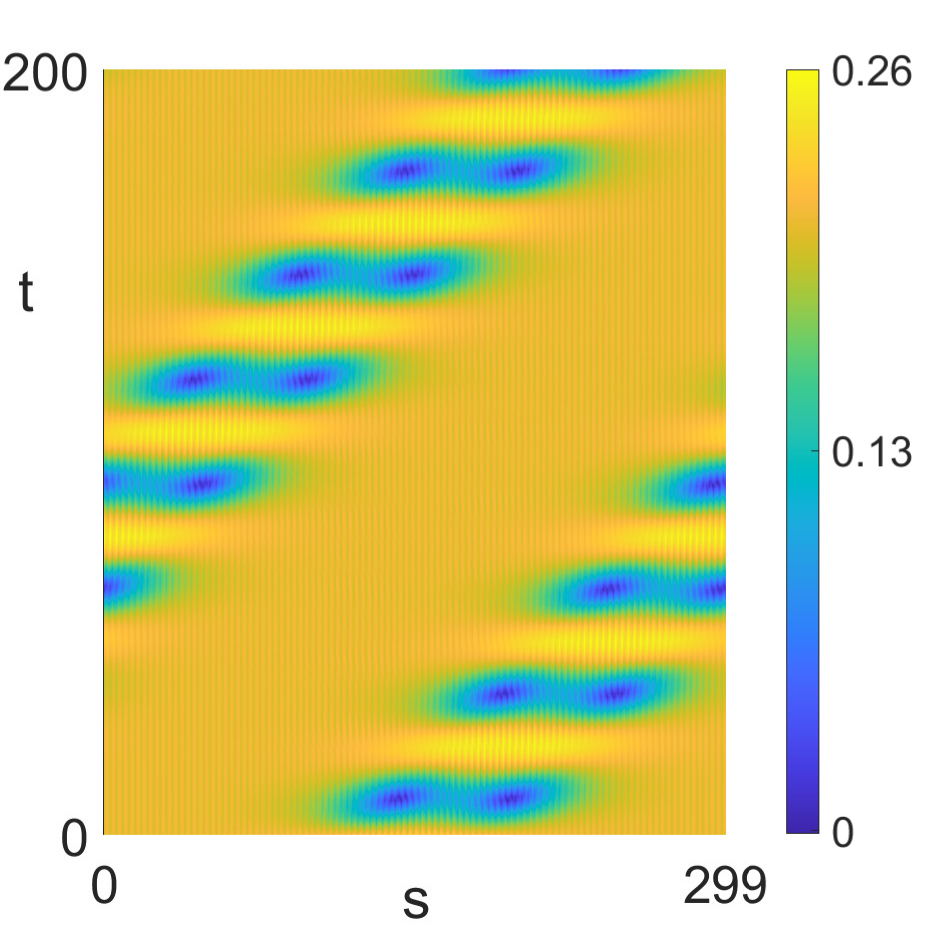}\label{fig:MTI2_BD_Left}}
\subfigure[]{\includegraphics[width=.23\textwidth]{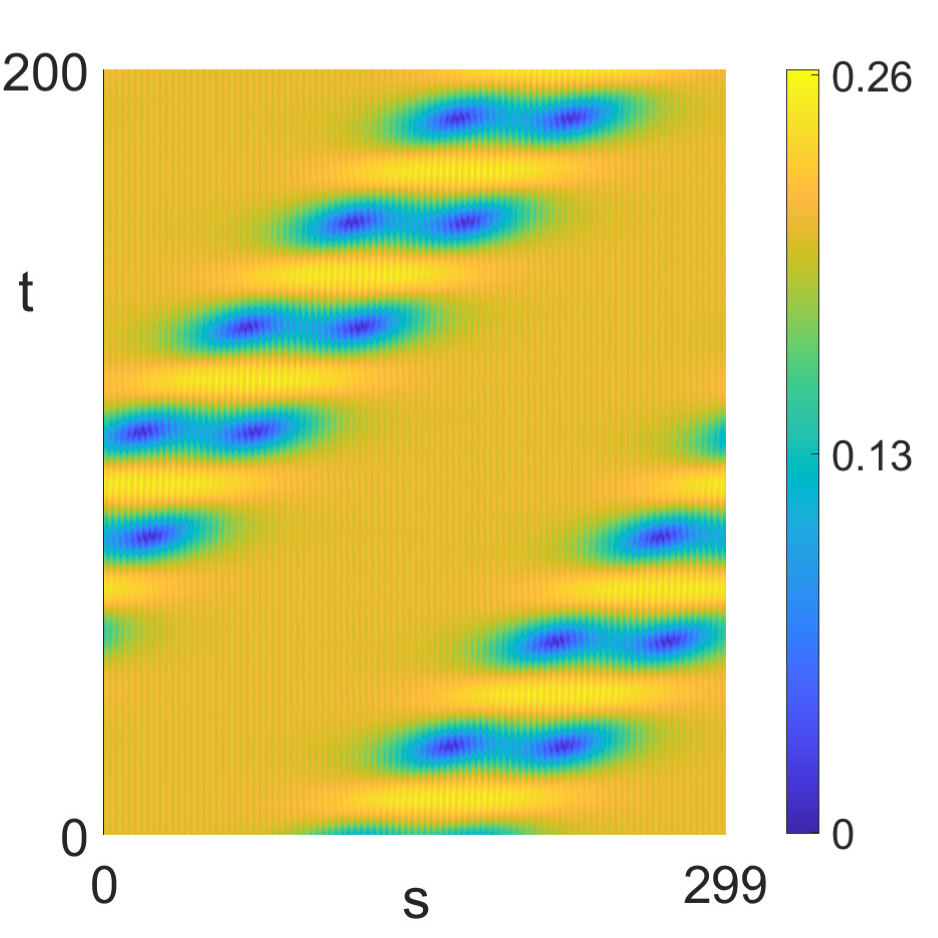}\label{fig:MTI2_BD_Right}}\\
\subfigure[]{\includegraphics[width=.23\textwidth]{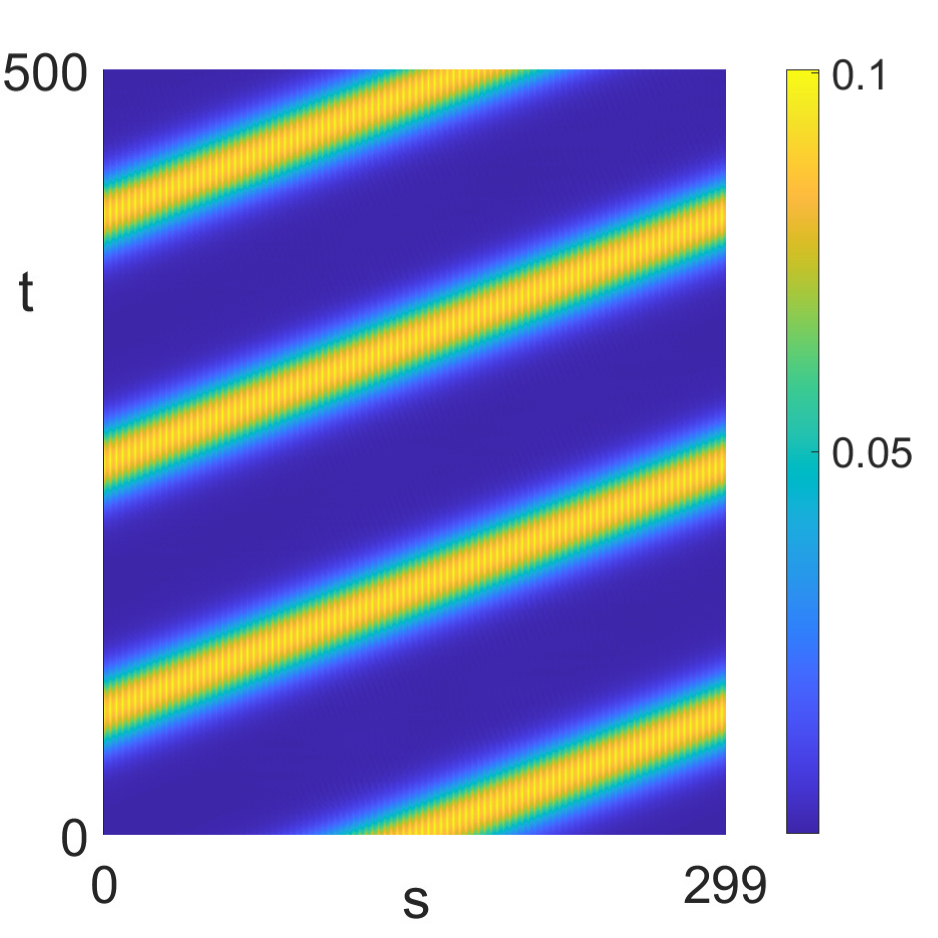}\label{fig:MTI2_BD_zA}}
\subfigure[]{\includegraphics[width=.23\textwidth]{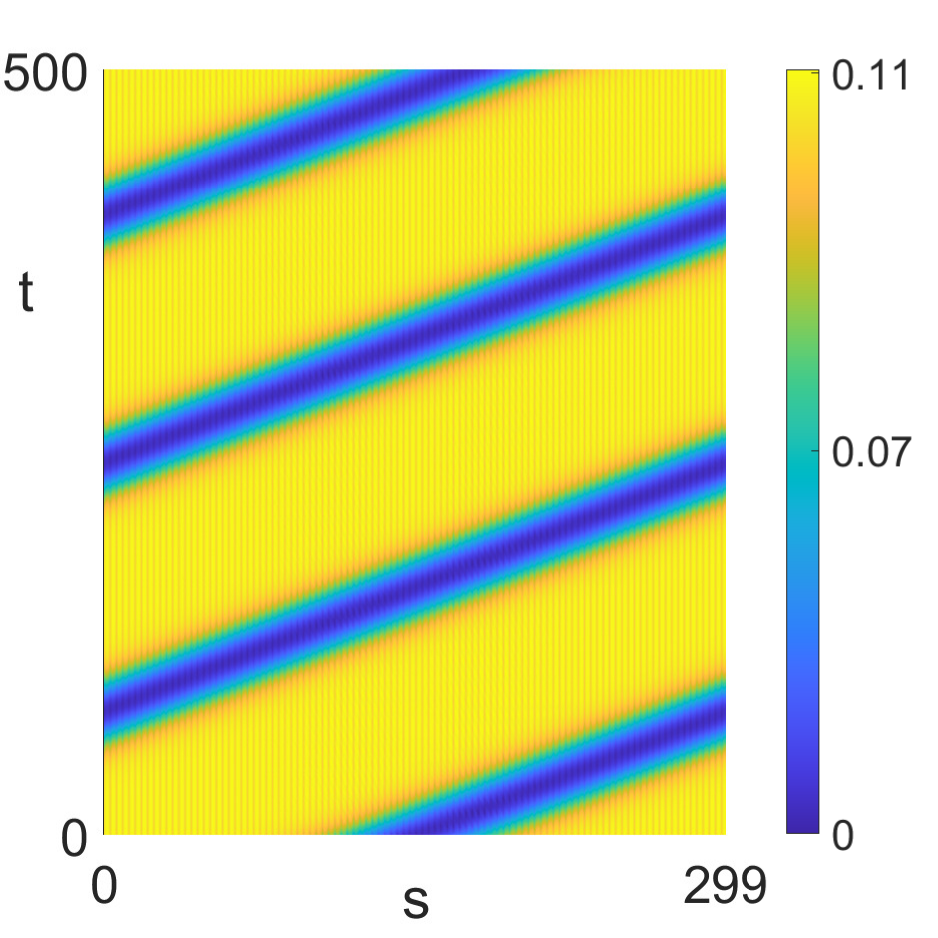}\label{fig:MTI2_BD_zB}}\\
\subfigure[]{\includegraphics[width=.23\textwidth]{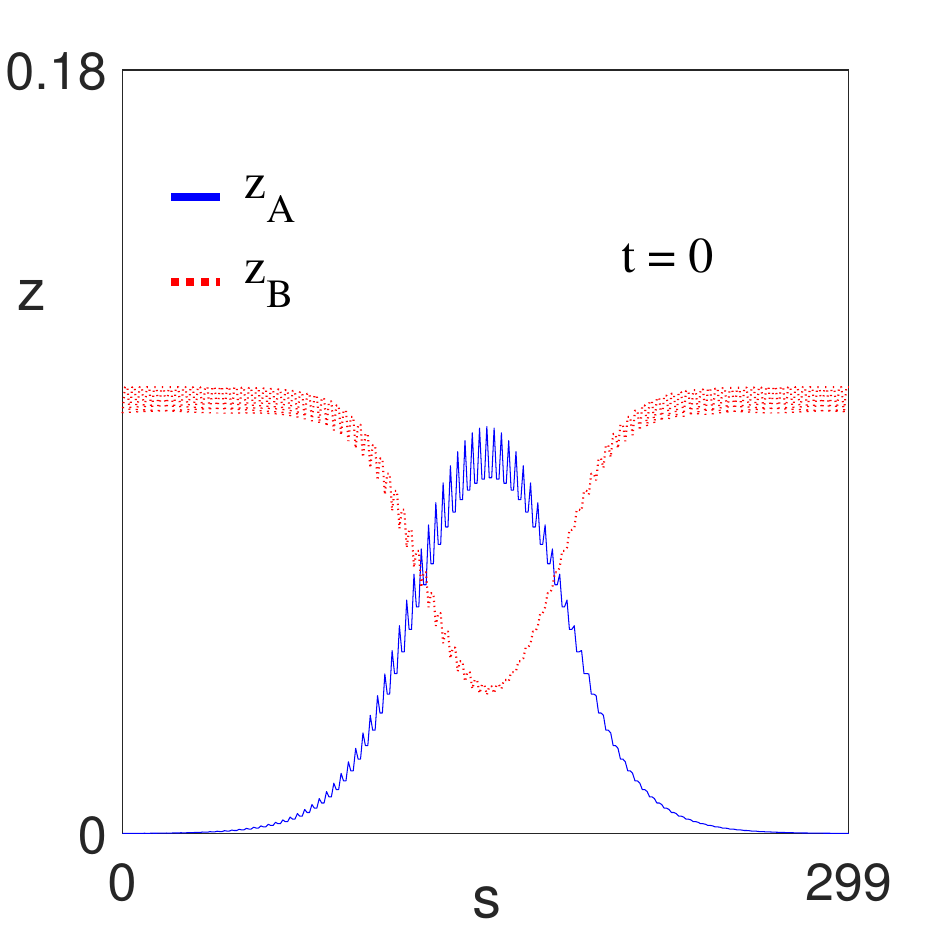}\label{fig:MTI2_BD_init}}
\subfigure[]{\includegraphics[width=.23\textwidth]{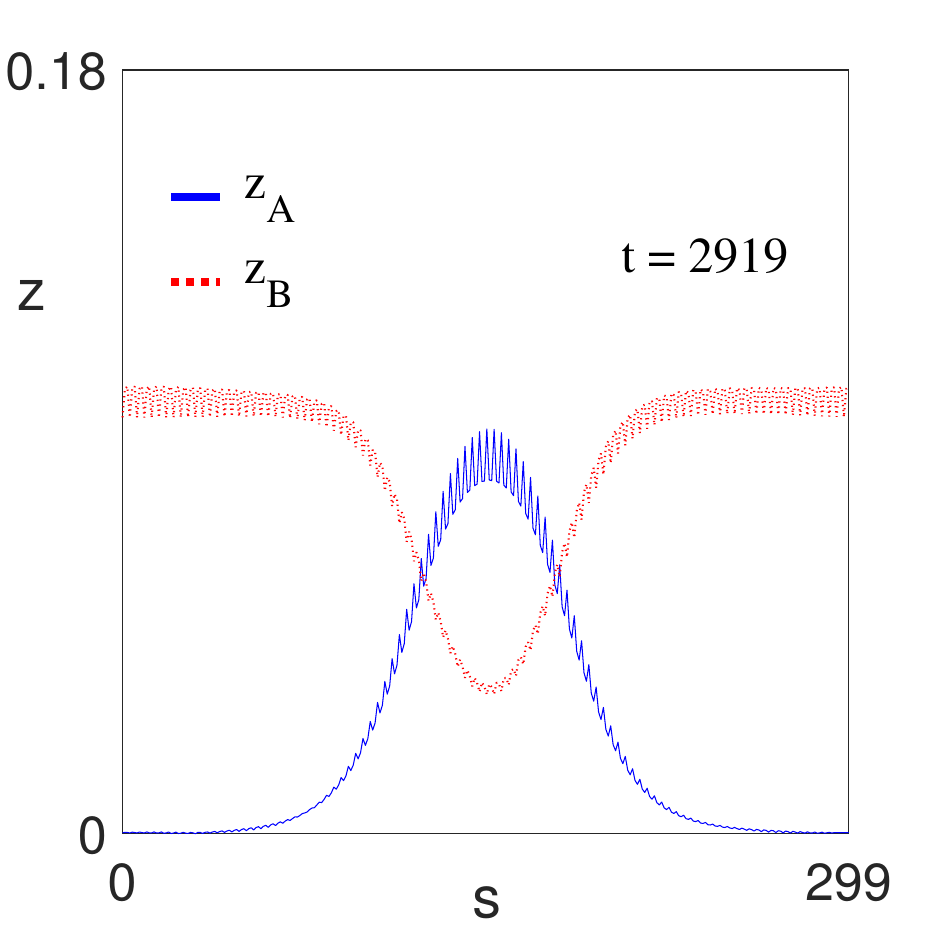}\label{fig:MTI2_BD_longtime}}
    \caption{(Color online) DB edge soliton propagating along the interface of the MTI with $N_r \times N_s=26 \times 300$ sites. (a,b) Space-time plots of the interface sites at $r=12$ and $r=13$ respectively. (c,d) Space-time plots of the excitation variables $z_\mathcal{A}$ and $z_\mathcal{B}$. (e,f) Profiles of the excitation variables $z_\mathcal{A}$ and $z_\mathcal{B}$ initially and after a long time.}
    \label{fig:BD_solution}
\end{figure}

Figure \ref{fig:BD_solution} shows a DB edge soliton propagating along the interface of the MTI. The excitation variables $z_\mathcal{A}$ and $z_\mathcal{B}$ simply exhibit the 1D CNLS dynamics as shown in Figs.~\ref{fig:MTI2_BD_zA} and \ref{fig:MTI2_BD_zB}. Comparison with Figs.~\ref{fig:MTI2_front_zA} and \ref{fig:MTI2_front_zB} shows that increasing $\omega_{\cal A}$ causes the two DWs to `merge' towards the center, which yields the DB soliton embedded in $ZN$. As shown in Figs.~\ref{fig:MTI2_BD_Left} and \ref{fig:MTI2_BD_Right}, in terms of the amplitude $z$ of the interface sites, a DB edge soliton resembles a dark breather in a stationary nonzero background, and the central beat has period $T_0$. The two dips forming this beat result from the `merger' of the two dips in Figs.~\ref{fig:MTI2_front_Left} and \ref{fig:MTI2_front_Right} towards the center.

This dip centered around zero on the space-time plots of site amplitudes is the defining feature of edge DWs and DB edge solitons at nearby parameters. Indeed, using Eq.~\eqref{eq:CM_ansatz}, the set of zeros on the $(\tilde{S},\tau)$-plane is given by
\begin{equation}\label{eq:curve-zeros}
    |{\cal A}(\tilde{S},\tau)\boldsymbol{X}_r^{(1)}|=|{\cal B}(\tilde{S},\tau)\boldsymbol{X}_r^{(2)}|.
\end{equation}
The zeros occur on this set with period $T_0$ at locations depending on the phase factors in Eq.~\eqref{eq:CM_ansatz}. If this set is a single line, then the two sides of Eq.~\eqref{eq:curve-zeros} intersect transversely as functions of $\tilde{S}$ for all $\tau$. Moreover, if this line separates two stationary nonzero backgrounds, then $({\cal A},{\cal B})$ tends to either $ZN$ or $NZ$ as $\tilde{S}\to\pm\infty$, so this solution must be a DW between $ZN$ and $NZ$.

Meanwhile, if the amplitude of either the left or the right interface site is zero, then the excitation variables $z_{\cal A}=z_{\cal B}$ by definition. Thus, for the DB edge soliton, the two intersections between the profiles of $z_{\cal A}$ and $z_{\cal B}$ in Fig.~\ref{fig:MTI2_BD_init} are precisely the intersections between the two lines of zeros and $t=0$ in Fig.~\ref{fig:MTI2_BD_Left} or \ref{fig:MTI2_BD_Right}. These profiles persist for a long time as shown in Fig.~\ref{fig:MTI2_BD_longtime}.

If the continuation parameter $\omega_{\cal A}$ is increased further in 1D CNLS, then the DB soliton will approach a weakly nonlinear DB soliton near a pitchfork bifurcation \cite{snee2024domain}. Thus, there will be no intersections between the profiles of $z_{\cal A}$ and $z_{\cal B}$, and the DB edge soliton will have a single nonzero dip, but we do not show such solutions explicitly.

\section{Topological protection}\label{sec:tp-bb}

\begin{figure}
\centering
\subfigure[]{\includegraphics[width=.25\textwidth]{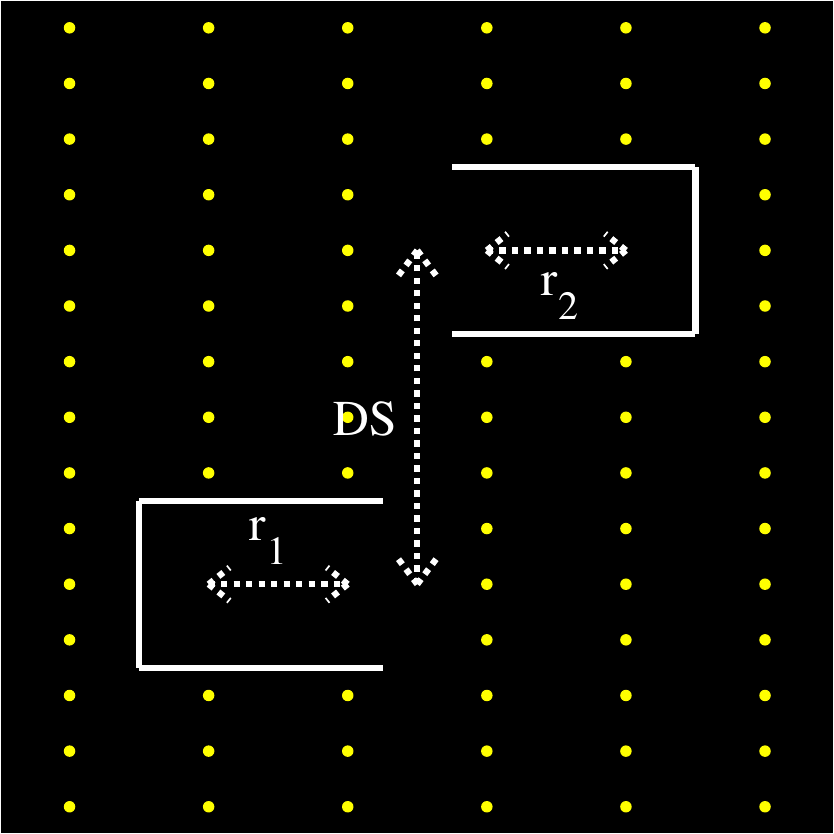}\label{fig:defect_diagram}}
\subfigure[]{\includegraphics[width=.45\textwidth]{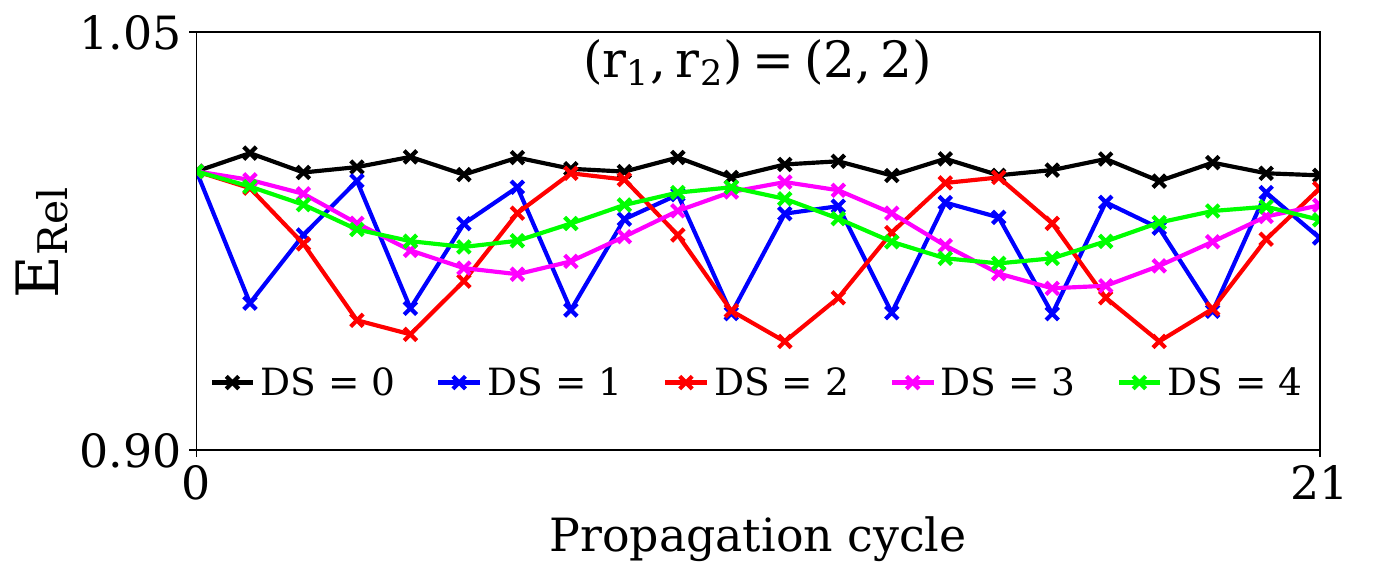}\label{fig:symmetry_scattering}}
\subfigure[]{\includegraphics[width=.45\textwidth]{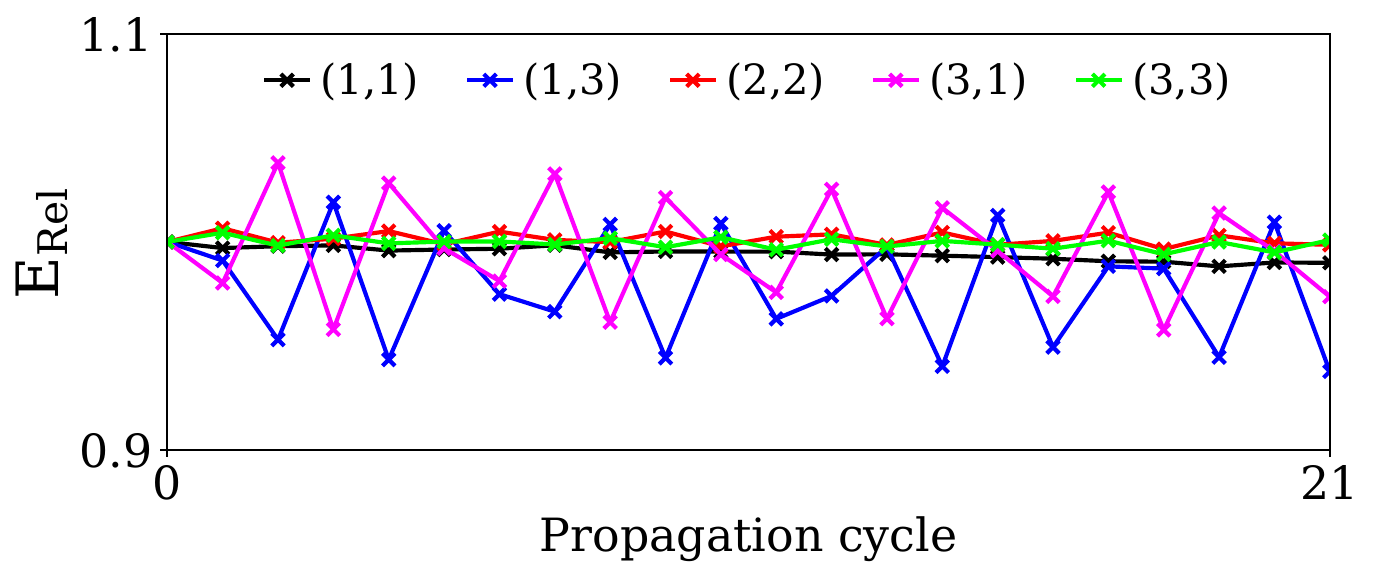}\label{fig:size_energy}}
    \caption{(Color online) (a) The parameter space $(r_1,r_2,\mbox{DS})$ of defects on the interface of the MTI. Here, each dot represents a site, and three vertical dots form a cell. (b,c) The relative energy $E_\text{Rel}$ of a BB edge soliton after each propagation cycle through defects with (b) $(r_1,r_2)=(2,2)$ and $\mbox{DS}\in[0,4]$; (c) $\mbox{DS}=0$ and $(r_1,r_2)$ in the legend.}
    \label{fig:symmetry_analysis}
\end{figure}

The topological protection of edge waves is often shown by passing them through boundary defects. For the MTI, we can build compact defects from the interface into either sub-lattice. Here, we carve a single defect into each sub-lattice from a single cell on the interface. We build the defect pair from interior cells only, so the periodic boundary condition in the $s$-direction remains. Thus, the parameter space of the defect pair is $(r_1,r_2,\mbox{DS})$, where $r_1$ and $r_2$ are the number of horizontal cells in the left and right sub-lattice respectively, and $\mbox{DS}$ is the distance in the $S$-direction between the two defects. Here, $|r_1-r_2|$ and $|\mbox{DS}|$ respectively quantify the horizontal and vertical asymmetry between the two defects. Figure \ref{fig:defect_diagram} shows an example of the defect pair with $(r_1,r_2,\mbox{DS})=(2,2,2)$.

We expect all nonlinear edge waves in Section \ref{sec:ves-dw} to be topologically protected when both carrier frequencies lie within a band gap. However, it is numerically hard to initialize a nonzero background on an irregular domain; see Ref.~\cite{snee2019edge} for a possible initialization of a scalar dark edge soliton. Thus, we consider a BB edge soliton as in Fig.~\ref{fig:BB_solution} but with the free parameters $(C_g,\gamma,\Theta_{\cal A},\Theta_{\cal B})=(0,-0.17,0,0)$ to better fit the soliton into the domain.

After each propagation cycle, we calculate the energy loss from the interface sites using the relative energy $E_\text{Rel}$ defined as the 2-norm of the excitation variables $(z_{\cal A},z_{\cal B})$ relative to the initial value. This quantity measures the combined mass of ${\cal A}$ and ${\cal B}$ on the interface. The energy concentration is also relevant, but the $\infty$-norm of $(z_{\cal A},z_{\cal B})$ entails an unnatural comparison between the maximum amplitudes of ${\cal A}$ and ${\cal B}$, which we will not show explicitly.

We first study BB edge soliton propagation through horizontally symmetric defects by fixing $(r_1,r_2)=(2,2)$ and varying $\mbox{DS}$. Here, we assume $\mbox{DS}\geq0$; the $\mbox{DS}<0$ case should behave similarly. We also assume $\mbox{DS}\leq4$ to avoid significant overlap with the initial envelope.

The numerical results are shown in Fig.~\ref{fig:symmetry_scattering}, where we plot the relative energy $E_\text{Rel}$ over 21 propagation cycles. A key observation is that the defect pair with $\mbox{DS}=0$, which can be viewed as a single totally symmetric defect, performs the best. Thus, although the dynamics deviate from CNLS as the soliton passes through the defect, the two sub-lattices are equally affected, so the soliton can regain its shape after passing through the defect.

For vertically asymmetric defects with $\mbox{DS}>0$, the soliton passes through first the left defect and then the right defect, so the left half-soliton lags behind the right half-soliton during the distance $\mbox{DS}$. Figure \ref{fig:symmetry_scattering} shows that this lag effect always causes the relative energy $E_\text{Rel}$ to drop slightly before recovering, and the cycle repeats. Larger $\mbox{DS}$ tends to preserve energy and prolong recovery, but the precise relations are not obvious.

Next, we study BB edge soliton propagation through vertically symmetric defects by fixing $\mbox{DS}=0$ and varying $(r_1,r_2)$. The numerical results are shown in Fig.~\ref{fig:size_energy}. A key observation is that the totally symmetric defects with $r_1=r_2$ perform the best, and the energy loss is negligible for $r_1=r_2=1$, 2, or 3, so size does not matter.

For horizontally asymmetric defects with $r_1\neq r_2$, the half-soliton passing through the longer defect lags behind the half-soliton passing through the shorter defect for each propagation cycle. Figure \ref{fig:size_energy} shows that this lag effect causes the relative energy $E_\text{Rel}$ to drop slightly before overshooting, and the cycle repeats. The energy fluctuations for the $(1,3)$ and $(3,1)$ defects are almost out of phase. In the long run, the $(3,1)$ defect appears to maintain the energy better, but the $(1,3)$ defect appears to maintain the amplitude better; the latter is harder to define as noted earlier and thus not shown explicitly.

Overall, the energy loss over many propagation cycles is small for the topologically protected BB edge soliton even if the defect pair is asymmetric. By contrast, we have checked that the energy loss is already large over few propagation cycles for a non-topologically protected BB edge soliton even if the defect pair is totally symmetric, but we will not show this simulation explicitly.

It would be interesting to develop a scattering theory of edge solitons around compact defects. Since Fig.~\ref{fig:symmetry_analysis} shows that symmetry matters more than size, we should study defects with a fixed size but different locations and shapes. Besides, we should study a connected defect as the fundamental unit before tackling disconnected defects as in Fig.~\ref{fig:symmetry_scattering}. Thus, it is natural to study a polyomino defect, i.e., a finite and connected subset of the square lattice \cite{golomb1996polyominoes}. For example, the $(1,3)$, $(2,2)$, and $(3,1)$ defects in Fig.~\ref{fig:size_energy} are a polyomino with four cells, i.e., a tetromino, and their locations relative to the interface in the $r$-direction determine their scattering properties.

Generally, the scattering properties of polyominoes depend on the spectra of the associated linear operators. Two linear operators are called isospectral if they have the same spectrum. Thus, one can ask to what extent the spectrum determines the location and shape of the polyomino. This generalizes a fundamental problem in spectral geometry of inferring the geometry of a domain from the spectrum of its Laplacian, which is well known as hearing the shape of a drum \cite{kac1966can,giraud2010hearing}. Our isospectral problem might be harder since the polyomino is a hole on the interface between two topological sectors.

\section{Vector edge soliton collisions}\label{sec:col-bb}
It is crucial to check that a vector soliton is not just scalar solitons acting independently. This vector nature is often shown by soliton collisions. In the defocusing CNLS equation, DB soliton collisions can exhibit mass exchange and modify the frequencies and phases of the bright components and the soliton velocities \cite{busch2001dark,snee2024domain}. The focusing Manakov system provides arguably the most prominent application, where BB soliton collisions can exhibit energy transfer described by linear fractional transformations and enable a universal Turing machine \cite{jakubowski1998state}. In the non-integrable regime, such collisions cannot be cascaded efficiently due to radiation, so collision-based computing entails restoration of full-energy solitons \cite{jakubowski2002computing}.

\begin{figure}
\centering
\subfigure[]{\includegraphics[width=.49\textwidth]{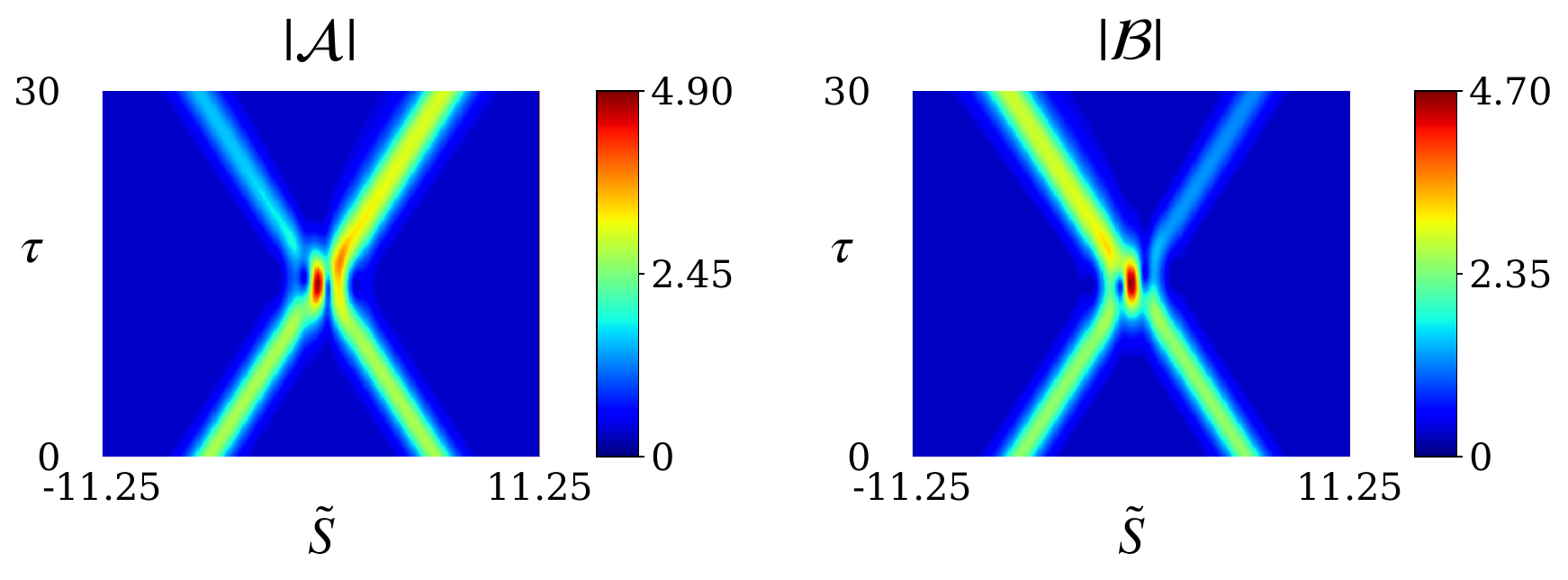}\label{fig:CNLS_BB_piover2}}
\subfigure[]{\includegraphics[width=.49\textwidth]{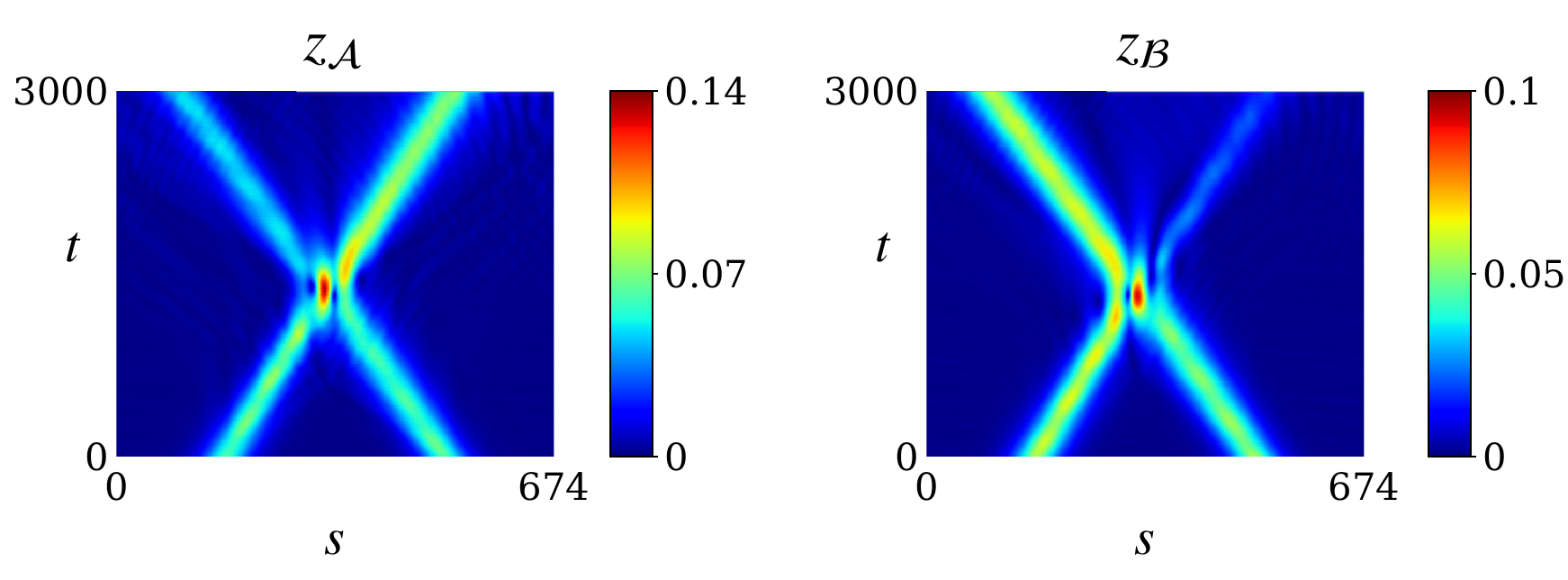}\label{fig:MTI_BB_piover2}}
    \caption{(Color online) Energy transfer in a collision between two BB (edge) solitons. (a) Space-time plots of $|{\cal A}|$ and $|{\cal B}|$ in the CNLS equation. (b) Space-time plots in the comoving frame of the excitation variables $z_\mathcal{A}$ and $z_\mathcal{B}$ in the MTI.}
    \label{fig:MTI_energy_transfer}
\end{figure}

Here, we realize a collision between two polarized BB solitons in the focusing CNLS equation in Section \ref{sec:focu-cnls}. Both solitons satisfy Eq.~(\ref{eq:BB_ansatz_A}) but with different phases, $(\Theta_\mathcal{A}^{(L)},\Theta_\mathcal{B}^{(L)})=(\pi/2,0)$ for the left soliton and $(\Theta_\mathcal{A}^{(R)},\Theta_\mathcal{B}^{(R)})=(0,0)$ for the right soliton. The remaining free parameters are chosen as $(C_g^{(L)},\gamma^{(L)})=(0.2,-0.6)$ and $(C_g^{(R)},\gamma^{(R)})=(-0.2,-0.6)$. As shown in Fig.~\ref{fig:CNLS_BB_piover2}, the collision with this phase choice yields maximal energy transfer that strengthens the ${\cal A}$ component of the right soliton and the ${\cal B}$ component of the left soliton.

Now we realize this collision in the MTI using the same spacetime domain, recalling that $\tilde{S}=\epsilon(S-V_gt)$ and $\tau=\epsilon^2t$ with $\epsilon=0.1$. The MTI dynamics are shown in Fig.~\ref{fig:MTI_BB_piover2} using the excitation variables $z_{\mathcal{A},\mathcal{B}}$ in the reference frame comoving with velocity $V_g$ in $S$, i.e., $3V_g$ in $s$. The energy transfer in the MTI dynamics indeed strengthens the $z_{\cal A}$ component of the right soliton and the $z_{\cal B}$ component of the left soliton as predicted by the CNLS dynamics, thus confirming the vector nature of BB edge solitons. However, the MTI dynamics exhibit more radiation than the CNLS dynamics likely due to higher-order terms becoming non-negligible over the simulation duration, which is well beyond $O(\epsilon^{-2})$.

The existence of energy transfer suggests a potential application of our MTI to collision-based computing. However, energy restoration becomes essential both after soliton collisions and during soliton propagation. Despite these challenges, unconventional computing in nonlinear MTIs can form a promising direction in the emerging field of mechanical computing \cite{yasuda2021mechanical}.

\section{Discussion}\label{sec:disc}
In this paper, we have derived a 1D two-component CNLS equation from a 2D MTI with on-site cubic nonlinearity. The CNLS equation is the universal envelope equation for the nonlinear interaction between two edge modes satisfying the EGV condition. Thus, the CNLS equation applies to other MTIs in the same symmetry class \cite{shah2024colloquium}. We have also formulated additional conditions for the integrability of the CNLS equation.

Next, we have designed the MTI with two topological sectors separated by an interface. In the uncoupled limit, the dispersion relation exhibits two-fold degeneracy. The interface couplings provide a large parameter space to find EGV points that enable favorable CNLS coefficients. Our design maintains the simplicity of the components in the original design \cite{Susstrunk-Huber-2015}, which is a key advantage of MTIs over photonic TIs and enables our solutions to be observed in tabletop experiments.

Numerically, we have found that EGV points in our MTI can enable either focusing or defocusing CNLS coefficients. Thus, we have realized vector edge solitons and DWs propagating along the interface, including BB edge solitons in the focusing case, and DD edge solitons, edge DWs, and DB edge solitons in the defocusing case. In terms of the site amplitudes, these solutions resemble bright and dark breathers. It would be interesting to check if such solutions can be numerically continued into true breathers in the strongly nonlinear regime.

We have explicitly shown topological protection of BB edge solitons by passing them through compact defects on the interface. We have also shown energy transfer in BB edge soliton collisions with potential application to collision-based computing. This unique combination between the topological and vector nature makes BB edge solitons ideal for fault-tolerant computing.

Nonlinear edge waves including scalar edge solitons in Ref.~\cite{snee2019edge} and vector edge solitons in this paper are found in 2D MTIs with on-site cubic nonlinearity. It would be interesting to find such waves in 2D MTIs with inter-site cubic nonlinearity \cite{Pal-etal-2018} or with nonlinearity-induced topological transition \cite{darabi2019tunable}. It is also interesting to explore the recently proposed chiral nonlinearity \cite{guo2025practical}, which can yield nonlinear edge waves without frequency shift.

Prototypical MTIs are discrete models, but one can design perturbative metamaterials from such models \cite{matlack2018designing}. This design principle applies to both acoustic \cite{xue2022topological} and elastic \cite{miniaci2021design} topological metamaterials. For such phononic materials, diverse sources of nonlinearity including material, geometric, and contact nonlinearities can enable rich nonlinear wave responses \cite{patil2022review}.

In the classification scheme of topological phonons \cite{susstrunk2016_PNAS}, our MTI is a high-frequency reciprocal metamaterial in symmetry class AII in 2D. An interesting challenge is to explore high-frequency nonreciprocal metamaterials in symmetry class A in 2D. The nonreciprocal elements used in such materials are typically gyroscopes \cite{NKRVTI_PNAS_2015,wang2015topological}, which can generate nonlinear edge waves due to the quadratic nonlinearity in Euler's rotation equations.

There is much recent interest in higher-order TIs \cite{schindler2018higher}. A $d$-dimensional $n$-th-order TI has $(d-n)$-dimensional topologically protected boundary states, while higher-dimensional states remain insulating. Thus, 2D second-order nonlinear TIs have stationary nonlinear corner states \cite{yi2024delocalization,prabith2024nonlinear}. A direct analog of our 2D first-order nonlinear TI is a 3D second-order nonlinear TI, which has traveling hinge solitons \cite{tao2020hinge}. An interesting direction is to construct vector hinge solitons, which may involve more than two sub-lattices depending on the dihedral angle between the two surfaces meeting at the hinge.

Our analysis of the MTI also relates to topolectrical circuits \cite{lee2018topolectrical,imhof2018topolectrical,kotwal2021active}. A conservative topolectrical circuit consists of capacitors and inductors only, so its linearized equation of motion is a second-order ODE defined by a dynamical matrix like that of MTIs. However, nonlinearity can differ between topolectrical circuits and MTIs.

Topological mechanics has enabled new designs of acoustic devices, e.g., robust sound manipulation using Helmholtz resonators \cite{yves2017topological} and acoustic delay lines with a reconfigurable geometry \cite{zhang2018topological}. For such applications, nonlinearity can prevent the dispersion of the signals via the formation of scalar edge solitons. Besides, nonlinearity can transform the signals via collisions between two vector edge solitons, which can suppress or amplify either component of either soliton. The ability to carry and transform information in the parameter space can also make edge solitons useful for emerging applications of topological mechanics to machine learning \cite{li2024training,li2026topological}.

\section*{Acknowledgments}
For the purpose of open access, the author has applied a Creative Commons Attribution (CC BY) licence to any Author Accepted Manuscript (AAM) version arising from this submission.

\appendix
\section{Localized solutions to the coupled nonlinear Schr\"odinger equation}\label{app:cnls}
The CNLS equation \eqref{eq:MTI_CNLS_EQN1} for the MTI is a special case of the CNLS equation with general coefficients \cite{snee2024domain}:
\begin{equation}\label{Eq:GCNLS}
    \begin{aligned}
        iA_{t} + d_1A_{xx} + A(g_1|A|^2 +  g_2|B|^2) = 0, \\
        iB_{t} + d_2B_{xx} + B(g_3|B|^2 +  g_4|A|^2)  = 0,
    \end{aligned}
\end{equation}
where $A$ and $B$ are complex fields, and $d_{1-2}$ and $g_{1-4}$ are real constants. Here, we review the general program to classify localized solutions to Eq.~(\ref{Eq:GCNLS}) including DWs and vector solitons \cite{snee2024domain}.

First, we apply the following traveling wave reduction:
\begin{equation*}
\begin{aligned}
&A=\phi(\zeta)\exp(i(k_Ax-\omega_A t)),\\
&B=\psi(\zeta)\exp(i(k_Bx-\omega_B t)),
\end{aligned}
\end{equation*}
where $\phi,\psi\in\mathbb{C}$ depend only on the comoving variable $\zeta=x-C_gt$, with $C_g \in \mathbb{R}$ being the group velocity, and $k_A,k_B\in\mathbb{R}$ and $\omega_A,\omega_B\in\mathbb{R}$ are the respective wavenumbers and frequencies. This reduces Eq.~\eqref{Eq:GCNLS} to a complex ODE in $\zeta$:
\begin{equation}\label{eq:complex-ode}
    \begin{aligned}
        &d_1\phi'' + i(2d_1k_A-C_g)\phi' + (\omega_A-d_1k_A^2)\phi\\
        &+ g_1|\phi|^2\phi +  g_2|\psi|^2\phi = 0,  \\
        &d_2\psi'' + i(2d_2k_B-C_g)\psi' + (\omega_B-d_2k_B^2)\psi\\
        &+ g_3|\psi|^2\psi +  g_4|\phi|^2\psi = 0.
    \end{aligned}
\end{equation}
Thus, plane waves, DWs, and vector solitons in Eq.~\eqref{Eq:GCNLS} become respectively equilibria, heteroclinic orbits, and homoclinic orbits in Eq.~\eqref{eq:complex-ode}.

We first seek heteroclinic and homoclinic orbits in Eq.~\eqref{eq:complex-ode} with the restriction
\begin{equation}\label{eq:cnls-res}
    C_g=2d_1k_A=2d_2k_B,
\end{equation}
and then extend them to the original Eq.~\eqref{eq:complex-ode} without this restriction. This yields an ODE with real coefficients:
\begin{equation}\label{eq:real-ode-com}
    \begin{aligned}
        &d_1\phi'' + \bomega_A\phi + g_1|\phi|^2\phi +  g_2|\psi|^2\phi = 0,\\
        &d_2\psi'' + \bomega_B\psi + g_3|\psi|^2\psi +  g_4|\phi|^2\psi = 0,
    \end{aligned}
\end{equation}
where
\begin{equation}\label{eq:bomega}
\bomega_A=\omega_A-\frac{C_g^2}{4d_1},\quad
\bomega_B=\omega_B-\frac{C_g^2}{4d_2}.
\end{equation}
Thus, Eq.~\eqref{eq:real-ode-com} is invariant under an ODE version of the Galilean transformation: $(C_g,\omega_A,\omega_B)\rightarrow(0,\bomega_A,\bomega_B)$. A further restriction to $\phi,\psi\in\mathbb{R}$ yields the real ODE:
\begin{equation}\label{eq:real-ode}
    \begin{aligned}
        &d_1\phi'' + \bomega_A\phi + g_1\phi^3 +  g_2\psi^2\phi = 0,\\
        &d_2\psi'' + \bomega_B\psi + g_3\psi^3 +  g_4\phi^2\psi = 0,
    \end{aligned}
\end{equation}
but one can show that heteroclinic and homoclinic orbits can be sought in Eq.~\eqref{eq:real-ode} without loss of generality.

The equilibria of Eq.~\eqref{eq:real-ode} can be classified into four types: $ZZ$, $ZN$, $NZ$, and $NN$, with $Z \ (N)$ denoting zero (nonzero) values. A heteroclinic orbit between two equilibria generically exists at a single value of a bifurcation parameter known as the Maxwell point. One can show that such orbits can only form between $ZZ$ and $NN$ or between $ZN$ and $NZ$. The former are called asymmetric DWs and exist in unconventional parameter regimes of Eq.~\eqref{Eq:GCNLS}; see Ref.~\cite{snee2024domain} for an example. The latter should then be called symmetric DWs, but they are often simply called DWs in the literature since they are common in the defocusing regime $d_1g_1<0$ and $d_2g_3<0$ when the immiscibility condition $d_1d_2(g_2g_4-g_1g_3)>0$ holds.

A DW and its symmetric partner can be assembled back-to-back to form a vector soliton; this construction is also called a heteroclinic cycle. Two asymmetric DWs between $ZZ$ and $NN$ can form an extended BB soliton with $NN$ embedded in $ZZ$ or an extended DD soliton with $ZZ$ embedded in $NN$. Two symmetric DWs between $ZN$ and $NZ$ can form extended DB solitons with $NZ$ embedded in $ZN$ or with $ZN$ embedded in $NZ$.

Moreover, two types of codimension-1 local bifurcations can yield vector solitons in Eq.~\eqref{eq:real-ode}: the subcritical Turing (Hamiltonian-Hopf) bifurcation and the pitchfork bifurcation. The former creates Turing solitons homoclinic to the $NN$ equilibrium with oscillatory tails and requires unconventional parameter regimes of Eq.~\eqref{Eq:GCNLS}; see Ref.~\cite{snee2024domain} for an example. The latter describes the bifurcation of $ZN$ or $NZ$ from $ZZ$ or the bifurcation of $NN$ from $ZN$ or $NZ$. The first case simply yields the well-known scalar bright or dark solitons in the scalar NLS equation, while the second case can yield DB, bright-antidark (BAD), DD, or dark-antidark (DAD) solitons, where antidark means a hump in a nonzero background. Note that BAD and DAD solitons were originally found in the BEC setting \cite{kevrekidis2004families}; these solutions do not exist in the symmetric CNLS equation in the optics setting.

Finally, exact solutions to the scalar NLS equation including scalar bright and dark solitons can be lifted to exact solutions to Eq.~\eqref{eq:real-ode} by requiring that a scalar soliton multiplied by a constant amplitude vector satisfies Eq.~\eqref{eq:real-ode}. This yields scalar BB and DD solitons when a consistency condition is satisfied.

The above three codimension-1 mechanisms to form vector solitons in Eq.~\eqref{eq:real-ode} are summarized in Table 1 in Ref.~\cite{snee2024domain}. Any such codimension-1 family of vector solitons can be numerically continued into a codimension-0 family of vector solitons in Eq.~\eqref{eq:real-ode}, i.e., a 2-parameter family in $(\bomega_A,\bomega_B)$ or equivalently a 3-parameter family in $(C_g,\omega_A,\omega_B)$ using the Galilean transformation.

Once we have this 3-parameter family in Eq.~\eqref{eq:real-ode}, we can remove the restriction \eqref{eq:cnls-res} and numerically continue these solutions in $(k_A,k_B)$ to obtain a 5-parameter family of complex homoclinic orbits in Eq.~\eqref{eq:complex-ode}, with the caveat that nontrivial continuation applies to a dark or antidark component, but not a bright component.

However, a bright component does provide an extra parameter for vector soliton collisions. In such collisions, a bright component can be non-trivially multiplied by a constant phase factor $e^{i\Theta_A}$ or $e^{i\Theta_B}$, while a dark or an antidark component cannot.

\bibliography{references.bib}

\end{document}